\documentclass[prd,twocolumn,showpacs,showkeys]{revtex4}
\usepackage{graphicx}
\usepackage{dcolumn}
\usepackage{bm}
\usepackage{amssymb}
\usepackage{comment}
\def\deltabar{{\mathchar'26\mkern-9mu\delta}}
\def\dbar{{\mathchar'26\mkern-12mu d}}
\begin{document}
\title{On the connection between Hamilton and Lagrange formalism\\ in Quantum
Field Theory}
\author{Selym \surname{Villalba-Ch\'avez$^1$}}
\author{Reinhard \surname{ Alkofer$^1$}}
\author{Kai \surname{Schwenzer$^{1,2}$}}
\affiliation{$^1$Institute of Physics, Universit\"atsplatz 5, A-8010 Graz, Austria.}
\affiliation{$^2$Department of Physics, Washington University, St. Louis, MO 63130, USA.}

\begin{abstract}
The connection between the Hamilton and the standard Lagrange formalism is
established for a generic Quantum Field Theory with vanishing vacuum expectation
values of the fundamental fields. The Effective Actions in both formalisms are the
same if and only if the fundamental fields and the momentum  fields  are related by
the stationarity condition. These momentum fields in general differ from the
canonical fields as defined via the Effective Action. By means of functional
methods a systematic procedure is presented to identify the full correlation 
functions, which depend on the momentum fields, as  functionals of those usually
appearing in the standard Lagrange formalism. Whereas Lagrange correlation
functions can be decomposed into tree diagrams the decomposition of Hamilton
correlation functions involves loop corrections  similar to those arising in
$n$-particle effective actions. To demonstrate the method we derive for theories
with linearized interactions the propagators of composite auxiliary  fields and the
ones of the fundamental degrees of freedom. The formalism is then utilized in the
case of Coulomb gauge Yang-Mills theory for which the relations between the   
two-point correlation functions of  the transversal and longitudinal  components 
of the conjugate momentum to the ones of the gauge field are given.   
\end{abstract}



\maketitle

\section{Introduction}

The path integral method within the Lagrange formalism is a fundamental tool to
formulate Quantum Field Theory. Certainly, the treatment of the same theory
within the canonical operator formalism is somewhat more cumbersome. In fact,
additionally to the operator-ordering problems of gauge theories
\cite{Schwinger:1962wd,Christ:1980ku}, the formal invariance properties of a
generic theory admit additional transformation laws which are not present in
the Lagrange formulation. The ``momentum fields'' in such a case are not related to
the fundamental fields by means of the canonical equations, see, {\it e.g.\/}, ref.\
\cite{Weinberg:1995mt},  but constitute separate degrees of freedom with
independent transformation properties and thus symmetries. As a consequence the
number of variables in the path integral increases in comparison to the
Lagrange formalism. This introduces for most theories considerable additional
complications in perturbative as well as non-perturbative calculations
\cite{Watson:2006yq,Watson:2007mz,Andrasi:2003zf,Andrasi:2005xu}.

On the other hand, some recent studies indicate that the first order formalism
proves to be a successful tool to study the required complete cancellation of the
energy divergences \cite{Zwanziger:1998ez} that emerge in a perturbative treatment
of Coulomb gauge Yang-Mills theory within the standard Lagrange formalism
\cite{Mohapatra:1971ue,Andrasi:2005xu}. Coulomb gauge Yang-Mills theory has
attracted attention since one possible solution to the confinement problem in QCD
is provided by the ``Gribov-Zwanziger'' scenario
\cite{Gribov:1977wm,Zwanziger:1998ez} ({\it cf.\/} also ref.\
\cite{Alkofer:2006fu}). However, the problem of renormalizing Coulomb gauge
Yang-Mills theory is still unsolved since these energy divergences cannot be
regularized using any of the standard procedures. Recently it was illustrated how
these energy divergences cancel at each order in perturbation theory
\cite{Niegawa:2006hg}. In order to perform explicit calculations a number of
methods have been applied as {\it e.g.\/} introducing a novel method to regularize
Feynman integrals in non-covariant gauges
\cite{Leibbrandt:1996tn,Heinrich:1999ak},  employing algebraic renormalizability
\cite{Zwanziger:1998ez}, and trying to recover Coulomb gauge Yang-Mills theory as a
limit of an interpolating gauge \cite{Baulieu:1998kx}. On the other hand, there are
several indications that the canonical or first order formalism  is better suited
for studying Coulomb gauge Yang-Mills theory
\cite{Watson:2006yq,Watson:2007mz,Andrasi:2003zf,Andrasi:2005xu,
Zwanziger:1998ez,Baulieu:1998kx,Szczepaniak:2001rg,Andrasi:2007dk,Watson:2007vc}.
Yet, even if such an approach would be successful, due to the dramatically
complicated form of the functional equations in the first order formalism, an
explicit non-perturbative study is far too involved to be computationally feasible.
Therefore, an analysis in the second order formalism would be highly desirable. To
this end we provide general connections between the Greens functions in the two
formulations that should help to perform the renormalization in the Lagrange
framework according to the insight in the renormalization procedure obtained in the
Hamilton framework.

In order to establish general relations between dressed correlation functions in 
the different formulations we exploit the following properties.
At vanishing sources associated to the momentum fields the Effective Action
({\it i.e.\/} the Generating Functional of one-particle-irreducible (1PI) Green's
functions) of the Hamilton formalism  reduces to the one of the Lagrangean
approach. This allows to reduce the set of Dyson-Schwinger equations (DSEs)
\cite{Dyson:1949ha,Schwinger:1951ex}
in the first order formalism to the corresponding set derived from the standard
path integral representation. An important special case is given if the
Hamiltonian is quadratic  in the momentum fields. For a corresponding Generating
Functional the integration over the momentum fields can be performed and
any $m-$point function involving this field  as the average of a polylocal
function of the quantum canonical momentum fields can be determined.
This means that the full correlation functions involving these canonical
variables can be found as a functional of those that usually appear in the
standard path integral representation.
The procedure to find these connections is closely related to
the functional method used in the derivation of the DSEs
(see {\it e.g.\/} ref.\ \cite{Alkofer:2000wg}).
In this paper we give the explicit form of these relations for a general
four-dimensional renormalizable theory. This includes the case that the
interaction terms involve the time derivative of the fields.
In a final step we resolve the relations between the proper two-point functions
in both frames by considering the inverse of the matrix-valued propagators in
the Hamiltonian approach.

Moreover, we show that similar connections arise for theories where not the 
kinetic but the interaction part is linearized. Such a bosonization procedure is 
an important technique used in many parts of physics ranging from hadronic physics
to condensed matter systems. Our results explicitly verify, in agreement with other
approaches, that there is no double counting in bosonized theories, but instead a
given correlation function in the underlying theory is exactly given by the sum of
all possible contributions involving both the fundamental and the composite degrees
of freedom in the bosonized theory.

This paper is organized as follows: In Sect.\ II the functional equations for DSEs
and a Symmetry-Related  identity (like Ward-Green-Takahashi, resp.\ Slavnov-Taylor,
identities (STIs) of a gauge theory 
\cite{Ward:1950xp,Green:1953te,Takahashi:1957xn,Taylor:1971ff,Slavnov:1972fg}) in
the phase space formulation are given. In Sect.\ III we present the first order
DSEs of theories which are quadratic in the momentum fields. In  addition,  we
derive a method to determine the correlation functions  and the STIs including
momentum fields from the respective quantities that usually appear in the standard
Lagrange representation. Diagrammatic rules and the general decomposition of the
full  proper functions in the Lagrange formalism are given in Sect.\ IV, while in
Sect.\ V and  Sect.\ VI we detail the explicit decomposition of the connected and
proper two-point functions in the Hamilton framework, respectively.  In Sect.\ VII
we show that the formalism can also be utilized in the case composite of auxiliary
fields, and last but not least in Sect.\ VIII we apply the formalism to the case of
Coulomb gauge Yang-Mills theory.  In the last section we conclude while essential
steps of many calculations have been deferred to several appendices.

\section{Functional equations}

We start our analysis by considering  a generic Quantum Field Theory formulated
within the  first order formalism. The Hamiltonian density  
$\mathcal{H}\left(q_m(x),p_m(x)\right)$ depends on the fundamental fields $q_m(x)$ 
and the conjugate momentum fields $p_m(x)$, respectively. As will be discussed in
Sect.\ IV.D, for fermionic fields the relation between the two formalisms is
trivial. Therefore we will treat only the case of bosonic fields in this and the
next section.

In the case of gauge invariant theories we will suppose that the Hamiltonian
$\mathcal{H}$ includes  the additional terms that arise when the constraints
associated to such theories, like {\it e.g.} Gauss' law in Quantum Electrodynamics,
are localized. In this context each Lagrange multiplier necessary to impose the
constraints will be treated as a fundamental field.  Certainly the presence of
these terms leads to the main differences between gauge theories and the more
conventional Hamiltonian systems. In addition, in a fixed gauge, ghost fields
appear. The usual path integral representation of these Grassmannian variables  is
formulated within the first order formalism (see also Sect.\ IV.C). Thus, in the
case of a gauge  theory they will be analysed in an independent way (for details
see Sect.\ \ref{sec:coulomb}). In this section, to set up the problem, 
we will disregard for the moment these potential complications.

Let us suppose, in addition, that the fields $p_m(x)$  and   $q_m(x)$  are  
coupled to  a set of  classical sources given by  $J^p_{m}(x)$ and $J^q_{m}(x)$, 
respectively. Under such conditions the source dependent vacuum-to-vacuum 
transition amplitude between the asymptotic states 
$\vert \textrm{Vac}, \textrm{in}\rangle$ and $\vert \textrm{Vac}, 
\textrm{out}\rangle$ looks like
\begin{eqnarray}
\mathcal{Z}[J]&=&\langle\mathrm{Vac},\mathrm{out}|\mathrm{Vac}, \mathrm{in}\rangle_{J}\nonumber\\&=&
\int \mathcal{D}[q]\mathcal{D}[p]
\exp\left\{\frac{i}{\hbar}\left(I\left[q,p,J\right]+i\epsilon\textrm{-terms}\right)\right\}\;\;\;\label{gen}
\end{eqnarray}
where $[q]$  denotes the collection of all fundamental fields, 
whereas $[p]$ the corresponding momentum fields. 
For more details see
section $9.2$ of the Ref.\ \cite{Weinberg:1995mt}.
Here  the argument in the exponential has the structure
\[
I[q,p,J]=I_0\left[q,p\right]+\int d^4x J(x)\cdot\phi(x)
\]
where
\[
\phi(x)\equiv\left(\begin{array}{c}
  p_m(x) \\ q_m(x)
\end{array}\right)\ \  \mathrm{and} \ \ J(x)\equiv\left(\begin{array}{c}
  J^p_{m}(x) \\ J^q_{m}(x)
\end{array}\right),
\]
include both momentum and fundamental fields and sources.
The components of the sources in this notation are distinguished
by upper labels, whereas
\begin{eqnarray}
I_0\left[q,p\right]&=& 
\int_{-\infty}^\infty d\tau \Bigg( \int d^3\textbf{x}\Big(
p_m(x)\dot{q}_m(x) 
\nonumber \\
&& \qquad \qquad \qquad -\mathcal{H}\left(q(x),p(x)\right)\Big) \Bigg)
\label{acv} 
\end{eqnarray}
looks like the classical expression for the Hamiltonian action.
Note that the part concerning to the $\epsilon$ terms in Eq.
(\ref{gen}) have the function to produce the necessary
$i\epsilon^{\prime}\textrm{s}$ in the denominators of all
propagators such that the correct boundary conditions  of the fields
at asymptotic times, $q_m(\textbf{x},\pm\infty)$, are implemented 
(see again Ref.\ \cite{Weinberg:1995mt}).

The fact that $I_0[q,p]$  looks like the action expressed in terms of canonical variables
is somewhat misleading since the momentum fields $p_m(x)$
are independent variables and therefore {\bf not yet} related to the
fundamental fields $q_m(x)$ or their derivatives. In particular,
since the path integral is not saturated by its saddle point(s) they are not
constrained to obey the equations of motion of classical Hamiltonian dynamics
${\delta I_0}/{\delta \phi}=0$ where
\begin{equation}
\frac{\delta I_0}{\delta \phi}\equiv\left(\begin{array}{c}
\frac{\delta I_0}{\delta p_m}  \\
\\
\frac{\delta I_0}{\delta q_m}
\end{array}\right)
=\left(\begin{array}{c}
\dot{q}_m-\frac{\partial\mathcal{H}}{\partial p_m}  \\
\\
\dot{p}_m-
\nabla\cdot\frac{\partial \mathcal{H}}{\partial\left(\nabla q_m\right)}+
\frac{\partial\mathcal{H}}{\partial q_m}
\end{array}\right).
\label{me1}
\end{equation}

The path integrals, however, contain the information about these equations of
motion. Indeed, let us consider the  operator version  of Eq.\ (\ref{me1}).
Its vacuum expectation values in the presence of the external classical sources
can be expressed as
\begin{equation}
\overline{\frac{\delta I_0}{\delta \phi}}=
\frac{\int\mathcal{D}[q]\mathcal{D}[p]\frac{\hbar}{i}\frac{\delta}{\delta\phi}
\exp\left\{\frac{i}{\hbar}I\left[q,p,J\right]\right\}}
{\langle0_\mathrm{out}\vert0_\mathrm{in}\rangle_{J}}
-J.
\end{equation}
Assuming the absence of any boundary terms, the integral of such a functional 
derivative vanishes. Then by substituting
each of the elementary objects present in Eq.\ (\ref{me1}) by the derivative
with regard to the respective classical sources, the following functional
differential equation arises
\begin{eqnarray}
\left.\frac{\delta I_0}{\delta \phi}\right\vert_{\phi(x)\to\frac{\hbar}{i}
\frac{\delta}{\delta J(x)}}\mathcal{Z}[J]=-J(x)\mathcal{Z}[J].
\label{partida1}
\end{eqnarray}
It contains all equations of motions fulfilled by all Green's functions.

Introducing the Generating Functional of connected Green's functions,
$\mathcal{W}^H[J]=-i\ln\left(\mathcal{Z}[J]\right)$,  allows us to
rewrite  Eq.\ (\ref{partida1})  as
\begin{eqnarray}
\left.\frac{\delta I_0}{\delta \phi}\right\vert_{\phi(x)\to
\frac{\delta\mathcal{W}^H}{\delta J(x)}+\frac{\hbar}{i}\frac{\delta}{\delta
J(x)}}=-J(x)\label{Wsde1},
\end{eqnarray} where in the last step we made use
of the following identity
\begin{equation}
\mathfrak{F}\left(\frac{\delta}{\delta J}\right)\exp(\mathfrak{G}(J))=
\exp(\mathfrak{G}(J)) \mathfrak{F}\left(\frac{\delta \mathfrak{G}(J)}{\delta J}
+\frac{\delta}{\delta J}\right)\label{especial}
\end{equation}
for arbitrary functionals $\mathfrak{F}$ and $\mathfrak{G}$.
This set of equations constitutes the DSEs for
the connected Green's functions.

In the next step we introduce the vacuum expectation values of the momentum and the 
fundamental fields in the presence of sources
\begin{equation}
\bar{\phi}[J]=\frac{1}{i}\frac{\delta\mathcal{W}^H}{\delta J(x)}\equiv
\left(\begin{array}{c}
   \bar{p}[J]\\
   \\\bar{q}[J]
  \end{array}\right)=\left(\begin{array}{c}
   \frac{1}{i} \frac{\delta\mathcal{W}^H}{\delta J^p_{m}}\\
   \\ \frac{1}{i} \frac{\delta\mathcal{W}^H}{\delta J^q_{m}}
  \end{array}\right).
\end{equation}
We assume  that it is possible  to invert these relations such  that
the sources are expressed as functionals of the vacuum expectation values of
$\hat{\Pi}(x)$ and $\hat{\mathcal{Q}}(x)$. Hereby the replacements of variables
in  Eq.\ (\ref{Wsde1})  becomes
\begin{equation}
\phi(x)\to\bar{\phi}[J](x)+
\frac{\hbar}{i}\int d^4x^\prime\mathbb{D}[J](x,x^\prime)
\frac{\delta}{\delta \bar{\phi}(x^\prime)}\label{r11}
\end{equation}
where
\begin{equation}
\mathbb{D}[J](x,x^{\prime})\equiv\left(\begin{array}{cccc}
  \Delta_{ml}^{pp}[J](x,x^{\prime}) &  \Delta_{ml}^{pq}[J](x,x^{\prime}) \\
   &   \\
  \Delta_{ml}^{qp}[J](x,x^{\prime}) & \Delta_{ml}^{qq}[J](x,x^{\prime})
\end{array}\right)\label{cpro}
\end{equation}
are the source-dependent two-point Green's function where
\begin{equation}
\Delta^{pq}=\frac{1}{i}\frac{\delta\mathcal{W}^H\left[J\right]}
{\delta J^{p}(x)\delta J^{q}(x^\prime)}=\frac{\delta \bar{q}_l[J](x^\prime)}
{\delta J^{p}(x)}
\end{equation}
and $\Delta^{pp}[J],$ $\Delta^{qq}[J]$ and $\Delta^{qp}[J]$ are analogously
defined.

The Effective Action in the first order formalism
$\Gamma^H[\bar{q}_{m},\bar{p}_{m}]$ can be defined by the Legendre transform
of $\mathcal{W}^H[J]$ with respect to the associated  ``averaged''
fields of the theory
\begin{equation}
\Gamma^H[\bar{\phi}]\stackrel{def}{\equiv}\mathcal{W}^H[J]-
\int d^4x J(x)\cdot\bar{\phi}(x).\label{ea}
\end{equation}
Remember that the compact notation implies a twofold Legendre transformation
in the two independent components $q$ and $p$.
By considering the functional derivative of $\Gamma^H[\bar{q},\bar{p}]$
with respect to  $\bar{p}_i(x)$ and $\bar{q}_i(x)$  we obtain as expected
\begin{equation}
\frac{\delta\Gamma^H}{\delta \bar{\phi}(x)}=-J(x). \label{eq:ecmov}
\end{equation}

Substituting  Eq.\ (\ref{eq:ecmov}) into Eq.\ (\ref{Wsde1}), however,  we obtain 
with the replacement Eq. (\ref{r11}) the desired functional
differential equations for $\Gamma^H[\bar{q},\bar{p}],$
\begin{equation}
\frac{\delta\Gamma^H}{\delta \bar{\phi}(x)}=
\left.\frac{\delta I_0}{\delta \phi(x)}\right\vert_{\phi(x)\to
\bar{\phi}[J](x)+
\frac{\hbar}{i}\int d^4x^{\prime}\mathbb{D}[J](x,x^\prime)
\frac{\delta}{\delta \bar{\Phi}(x^\prime)}} \, .\label{sdev}
\end{equation}
This relation generates all DSEs for the 1PI Green's functions of the first order
formalism. The method to obtain such a functional relation will be employed several times
in the following.

The derivation of the proper propagators is straightforward when keeping in
mind that they are  embedded in a  matrix.  In fact, taking a
derivative with regard  $\phi(x^\prime)$ in Eq. (\ref{sdev}) we obtain
\begin{equation}
\mathbb{G[\phi]}(x^{\prime\prime},x^\prime)\equiv\left(\begin{array}{cccc}
\frac{\delta^2\Gamma^H}{\delta\bar{p_m}(x^{\prime\prime})
\delta\bar{p}_n(x^\prime)} & \frac{\delta^2\Gamma^H}
{\delta\bar{p_m}(x^{\prime\prime})\delta\bar{q}_n(x^\prime)} \\
& \\ \frac{\delta^2\Gamma^H}{\delta\bar{q_m}(x^{\prime\prime})
\delta\bar{p}_n(x^\prime)} & \frac{\delta^2\Gamma^H}
{\delta\bar{q_m}(x^{\prime\prime})\delta\bar{q}_n(x^\prime)}
\end{array}\right).\label{pgf}
\end{equation}
This expression  and Eq. (\ref{cpro}) are related via the
identity
\begin{equation}
\int d^4x^{\prime\prime}\mathbb{D}[J](x,x^{\prime\prime})
\mathbb{G[\phi]}(x^{\prime\prime},x^\prime)=-\mathbb{I},\label{unidad}
\end{equation}
which can be obtained by taking a derivative with respect to the source $J$ in
Eq.\ (\ref{eq:ecmov}). Hereby
$\mathbb{I}=\delta_{mn}\delta^4(x-x^\prime)\otimes 1_{2\times2}$ denotes the
identity in the considered space.
Any other connected or proper Green's function can be derived by considering
higher order derivatives with respect to $J$
and $\phi$ in  Eq.\ (\ref{cpro}) and  Eq.\ (\ref{pgf}), respectively.

In general, these  functions  are constrained by the symmetry
properties of the initial ``action''. In the case of a gauge theory a corresponding 
derivation leads
to the  STIs. For Coulomb gauge Yang-Mills theory it is presented  in
Sect. \ref{sec:coulomb}. Here we will illustrate the potential complications
arising in the first  order formalism. For this  
it sufficient to assume that $I_0$ is  
invariant under the simultaneous infinitesimal transformations
\begin{equation}
\delta \phi[\phi]\equiv\epsilon \cdot \left(\begin{array}{c}
\mathfrak{G}_m^q\left[p,q\right] \\
\\ \mathfrak{F}_m^p\left[p,q\right]\end{array}\right)\label{transfi}
\end{equation}
 and also assume that they leave the path integration measure invariant.
Performing this substitution in Eq.\ (\ref{gen}) we can obtain the following
``Symmetry-Related Functional Identity''
\begin{equation}
0=\int\mathcal{D}[q]\mathcal{D}[p]\int d^4 x J(x)\cdot\delta\phi(x)
\exp\left[\frac{i}{\hbar}I\left[\phi,J\right]\right].\label{slavnov}
\end{equation}
In the next step we substitute the fundamental and momentum fields by the 
respective derivatives with respect to the classical source. Following a 
procedure similar to the derivation of the DSEs we rewrite 
the symmetry-related identity Eq.\ (\ref{slavnov}) and express it  in terms of 
the full Generating Functional  $\mathcal{Z}[J]$:
\begin{equation}
0=\int d^4x J(x)\cdot\delta\phi\left[\frac{\hbar}{i}\frac{\delta}{\delta J}
\right]\mathcal{Z}[J].\label{comparar}
\end{equation}
Analogously, the above relation can be rewritten for connected Green's functions
\begin{equation}
0=\int d^4 x
J(x)\cdot\delta\phi[\phi]\vert_{\phi\to\frac{\delta\mathcal{W}^H}{\delta
J}+\frac{\hbar}{i}\frac{\delta}{\delta J}} .\label{comparar1}
\end{equation}
As before,  Eq.\ (\ref{slavnov}) can be reformulated in terms of the
Effective Action
\begin{equation}
0=\left.\int d^4x \frac{\delta\Gamma^H}{\delta
\bar{\phi}}\cdot\delta\phi[\phi]\right\vert_{\phi\to\bar{\phi}[J]+
\frac{\hbar}{i}\int
d^4y\mathbb{D}[J](x,y)\frac{\delta}{\delta \bar{\phi}(y)}} .\label{comparar2}
\end{equation}
This identity verifies that the Effective Action in first order formalism
preserves the continuous symmetries of the initial canonical quantum  action.

At this point the DSEs and the symmetry-related identity  Eq.\ (\ref{slavnov}) 
of a QFT seem to be more cumbersome than the usual one which appear in the
standard path integral formulation. However, in the next section we shall show
that under certain conditions  the above  system can be reduced to the latter
one.

\section{The connection between Lagrange and Hamilton formalism}

From here on we will denote all the variables a field depends on by a single
latin index. For instance in case of a gauge field such indices will indicate
the space-time point $x$, the vectorial index $\mu$, and the  adjoint
gauge group index $a$. Repeated indices are summed and integrated over for
discrete and continuous variables, respectively. Also all fundamental fields
will be denoted by the same letter $q$. Similarly, $p$ will represent all
momentum fields. 

Let us now consider the path integral representation of a
$n$-point function involving only momentum fields
\begin{eqnarray}
\Delta_{i_1\ldots i_n}^{p\ldots p}&=&
\frac{\langle0_\mathrm{out}\left\vert
T\left\{\hat{\Pi}_{i_1}\ldots\hat{\Pi}_{i_n}\right\}
\right\vert0_\mathrm{in}\rangle_{J}}
{i^{1-n}\langle0_\mathrm{out}\vert0_\mathrm{in}\rangle_{J}}
\label{Delta}\\
&=&\frac{\int\mathcal{D}[q]\mathcal{D}[p]p_{i_1}\ldots
p_{i_n}\exp\left[\frac{i}{\hbar}I[q,p,J]\right]}
{i^{1-n}\int\mathcal{D}[q]\mathcal{D}[p]
\exp\left[\frac{i}{\hbar}I[q,p,J]\right]}\;\;\; .\nonumber
\end{eqnarray}
Here $T$ is the time ordering operator, and 
$\hat{\Pi}_{i}\equiv\hat{\Pi}_m(\textbf{x},\tau)$ are the Hermitian momentum 
field operators in the Heisenberg picture. The expression (\ref{Delta})
can be written via functional derivatives with respect to $J^p_{i}$ as 
\begin{equation}
\Delta_{i_1\ldots i_n}^{p\ldots
p}=\frac{\int\mathcal{D}[q]\mathcal{D}[p]\frac{\hbar}{i}\frac{\delta}{\delta
J^p_{i_1}}\ldots\frac{\hbar}{i}\frac{\delta}{\delta
J^p_{i_n}}\exp\left[\frac{i}{\hbar}\left\{
I[q,p,J]\right\}\right]}{i^{1-n}\int\mathcal{D}[q]\mathcal{D}[p]
\exp\left[\frac{i}{\hbar}\left\{I[q,p,J]\right\}\right]}.\label{ass}
\end{equation}

In the following we will consider only the important class of Hamiltonians 
which are at most quadratic in the momentum fields. Note that this case is 
realized for all renormalizable and most non-renormalizable theories.
For later use we write the Hamiltonian in the form
\begin{equation}
\mathcal{H}=\frac{1}{2} p_i\mathcal{A}_{ij}[q]p_m+\mathcal{B}_i[q]p_i+
\mathcal{C}[q],\label{Hamiltoniand1}
\end{equation}
which defines a real, symmetric, positive and non-singular matrix 
$\mathcal{A}$ as well as the real functionals of $q$, $\mathcal{B}_i[q]$ and
$\mathcal{C}[q]$. For the considered class of Hamiltonians the Gaussian 
integration over $p$ can be performed analytically. As shown in appendix 
\ref{app:p-int} it yields an expression of the form
\begin{equation}
\Delta_{i_1\ldots i_n}^{p\ldots
p}=\frac{\int\mathcal{D}[q]\frac{\hbar}{i}\frac{\delta}{\delta
J^p_{i_1}}\ldots\frac{\hbar}{i}\frac{\delta}{\delta
J^p_{i_n}}\exp{\left[\frac{i}{\hbar}\tilde{\mathcal{S}}[q,J]\right]}}
{i^{1-n}\int
\mathcal{D}[q]\exp{\left[\frac{i}{\hbar}\tilde{\mathcal{S}}[q,J]\right]}}.
\label{ass111}
\end{equation}
The general result for the new action arising in the exponential is given in 
Eq.\ (\ref{gac}). 

For simplicity we will discuss first  the standard Hamiltonian with
$\mathcal{A}_{ij}[q]=\delta_{ij}$.  
The general case where $\mathcal{A}$ is  non-trivial is realized 
{\it e.g.} in the case of Coulomb gauge QCD which will be discussed in
Sect.\ \ref{sec:coulomb}.  (For the moment we only note that then the inverse 
matrix, $\mathcal{A}^{-1}[q]$, appears in the following expressions as a 
prefactor of $J_p$ making in general the action $S_0$ non-local.) 
In the considered simpler standard case the new action possesses the structure
\begin{equation}
\tilde{\mathcal{S}}[q,J]=\mathcal{S}_0+
\frac{1}{2}J^p_{i}J^p_{i}+J^p_{i}\frac{\delta\mathcal{S}_0}{\delta\dot{q}_i}+
J^q_{i}q_i
\label{actionLF}
\end{equation}
where $\mathcal{S}_0$ is the standard action
\begin{equation}
\mathcal{S}_0[q]=\frac{1}{2}\dot{q}^i \dot{q}^i
-\dot{q}^i \mathcal{B}^i[q]\ +\frac{1}{2}\mathcal{B}^i[q]\mathcal{B}^i[q]-
\mathcal{C}[q]  \, .\label{act}
\end{equation}

The application of  Eq.\ (\ref{especial}) on the integrand present in Eq.\
(\ref{ass111}) makes  it possible to write
\begin{equation}
\Delta_{i_1\ldots i_n}^{p\ldots
p}=\frac{\int\mathcal{D}[q]\left\{\hat{\mathcal{O}}_{i_1\ldots
i_{n-1}}^{p\ldots p} \left(J^p_{i_n}+\frac{\delta\mathcal{S}_0}{\delta
\dot{q}_{i_n}}\right)\right\}\exp{\left[\frac{i}{\hbar}\tilde{\mathcal{S}}
\right]}}{i^{1-n}\int
\mathcal{D}[q]\exp{\left[\frac{i}{\hbar}\tilde{\mathcal{S}}\right]}}.
\label{ass11}
\end{equation}
where the operator
\begin{equation}
\hat{\mathcal{O}}_{i_1\ldots i_{n-1}}^{p\ldots
p}=\prod_{l=1}^{n-1}\left( J^p_{i_l}+\frac{\delta\mathcal{S}_0}{\delta
\dot{q}_{i_l}}+\frac{\hbar}{i}\frac{\delta}{\delta
J^p_{i_l}}\right) \label{vass11}
\end{equation}
only acts on the function inside the curly brackets.
Note that the integrand in the numerator is a functional depending on the
fields $q$ as well as the sources $J_q$ and $J_p$.

\subsection{Alternative form of the DSEs}
Analogous to Eq.\ (\ref{Wsde1}) we can write the partially integrated
expression Eq.\ (\ref{ass11}) in terms of the Generating Functional of
connected Green's functions. Let us specialize to the case of one-point
Green's functions, \emph{i.e.,} the vacuum expectation values of the
momentum fields
\begin{equation}
\bar{p}_{i}[J]=\left.\frac{\delta\mathcal{S}_0}{\delta \dot{q}_i}
\right\vert_{q\to\frac{\delta\mathcal{W}^H}{\delta J^q}+
\frac{\hbar}{i}\frac{\delta}{\delta J^q}}+J^p_{i}=
\frac{\delta \mathcal{W}^H}{\delta J^p_{i}(\textbf{x},\tau)}.\label{mge1}
\end{equation}
 Eq.\ (\ref{eq:ecmov})  allows to  express
Eq.\ (\ref{mge1})  as
\begin{equation}
\frac{\delta\Gamma^H}{\delta\bar{p}_i}=
-\bar{p}_{i}+\left.\frac{\delta\mathcal{S}_0}
{\delta \dot{q}_i}\right\vert_{q\to\bar{q}[J]+\frac{\hbar}{i}
\Delta^{qq}[J]\frac{\delta}{\delta \bar{q}}}.\label{mge11}
\end{equation}
Here the dependence on the average momentum fields is made explicit and the action
in the Lagrange formalism appears which depends only on the $q$-fields.
From a functional point of view it is more convenient to express this via
Eq.\ (\ref{mge1})  as
\begin{equation}
\frac{\delta\Gamma^H}{\delta\bar{p}_i}=-\left[p_{i}-
\frac{\delta\mathcal{S}_0}{\delta \dot{q}_i}\right]_{\phi\to\bar{\phi}[J]+
\frac{\hbar}{i}\Delta^{\phi\phi}[J]\frac{\delta}{\delta \phi}}.\label{mge11v}
\end{equation}
Now,  employing  the identity
\begin{equation}
0=\int\mathcal{D}[q]\frac{\delta}{\delta q_m}\int\mathcal{D}[p]
\exp\left[\frac{i}{\hbar}I[q,p,J]\right]\label{ef1sd}
\end{equation}
we perform  the $p-$integration, so that the above expression reads
\begin{equation}
0=\int \mathcal{D}[q]\left(\frac{\delta\mathcal{S}_0}{\delta q_i}+
J^p_{m}\frac{\delta^2\mathcal{S}_0}{\delta q_i\delta \dot{q}_m}+J^q_{i}\right)
\exp\left[\frac{i}{\hbar}\tilde{\mathcal{S}}[q,J]\right],
\end{equation}
which makes it possible to obtain an analogous form of the field equations
\begin{equation}
\frac{\delta \Gamma^H}{\delta \bar{q}_i}=
\left.\left(\frac{\delta\mathcal{S}_0}{\delta q_i}-
\frac{\delta \Gamma^H}{\delta \bar{p}_m}
\frac{\delta^2\mathcal{S}_0}{\delta q_i\delta \dot{q}_m}\right)
\right\vert_{\phi\to \bar{\phi}[J]+\frac{\hbar}{i}\Delta^{\phi\phi}[J]
\frac{\delta}{\delta \bar{\phi}}}.\label{sdeqqq}
\end{equation}
Eqs.\ (\ref{mge11v}) and (\ref{sdeqqq}) represent alternative forms  of the
first order DSEs (\ref{sdev}).

This alternative form of the DSEs allows us
to find corresponding equations for the fundamental proper Green's functions
by taking derivatives with respect to the momentum and/or fundamental fields.
This way
the proper momentum propagator can be written as
\begin{equation}
\frac{\delta^2\Gamma^H}{\delta\bar{p}_i\delta\bar{p}_j}=
-\delta_{ij}+\frac{\delta}{\delta\bar{p}_i}
\left(\left.\frac{\delta\mathcal{S}_0}{\delta \dot{q}_j}
\right\vert_{q\to\bar{q}[J]+\frac{\hbar}{i}\Delta^{qq}[J]
\frac{\delta}{\delta \bar{q}}}\right),\label{mge12}
\end{equation}
whereas  the corresponding mixed Green's function is
\begin{equation}
\frac{\delta^2\Gamma^H}{\delta\bar{q}_i\delta\bar{p}_j}=
\frac{\delta}{\delta\bar{q}_i}\left(\left.\frac{\delta\mathcal{S}_0}
{\delta \dot{q}_j}\right\vert_{q\to\bar{q}[J]+\frac{\hbar}{i}\Delta^{qq}[J]
\frac{\delta}{\delta \bar{q}}}\right).\label{mge13}
\end{equation}

In a similar way, the other mixed propagator can be written as
\begin{eqnarray}
\frac{\delta^2 \Gamma^H}{\delta \bar{p}_i\delta \bar{q}_j}&=&
\frac{\delta}{\delta \bar{p}_i}
\left(\left.\frac{\delta\mathcal{S}_0}{\delta q_j}
\right\vert_{q\to \bar{q}[J]+\frac{\hbar}{i}\Delta^{qq}[J]\frac{\delta}
{\delta \bar{q}} }\right)\nonumber\\&-&\left.\frac{\delta^2\Gamma^H}
{\delta \bar{p}_i\delta\bar{p}_m}\frac{\delta^2\mathcal{S}_0}
{\delta q_j\delta \dot{q}_m}\right\vert_{\phi\to \bar{\phi}[J]
+\frac{\hbar}{i}\Delta^{\phi\phi}[J]\frac{\delta}{\delta \bar{\phi}} }
\label{sppdeqqqpr},
\end{eqnarray}
whereas the fundamental field propagator is given by
\begin{eqnarray}
\frac{\delta^2 \Gamma^H}{\delta \bar{q}_i\delta \bar{q}_j}&=&
\frac{\delta}{\delta \bar{q}_i}
\left(\left.\frac{\delta\mathcal{S}_0}{\delta q_j}\right\vert_{q\to \bar{q}[J]+
\frac{\hbar}{i}\Delta^{qq}[J]\frac{\delta}{\delta \bar{q}} }\right)\nonumber\\&-&
\left.\frac{\delta^2\Gamma^H}{\delta \bar{q}_i\delta\bar{p}_m}
\frac{\delta^2\mathcal{S}_0}{\delta q_j\delta \dot{q}_m}
\right\vert_{\phi_j\to \bar{\phi}[J]+\frac{\hbar}{i}\Delta^{\phi\phi}[J]
\frac{\delta}{\delta \bar{\phi}} }.\label{sppdeqqq}
\end{eqnarray}
Hereby, in the derivation of
Eqs.\ (\ref{sppdeqqqpr}) and (\ref{sppdeqqq}) we have discarded those terms 
that vanish when the sources are set to zero. In addition, note that
the proliferated occurrence of Green's functions involving the $p-$field
arises from functional derivatives
on $\Delta_{ij}^{qq}[J]$. As will become clear below, despite the different
appearance, the structure of Eqs.\ (\ref{sppdeqqqpr}) and (\ref{sppdeqqq})
is related to the one of Eqs.\ (\ref{mge12}) and (\ref{mge13}).

\subsection{Pure and mixed momentum correlation functions\label{gen-Ham-con}}

In this subsection we derive expressions for first order correlation functions
entirely in terms of second order correlation functions. To this end, we start
from Eq.\ (\ref{ass11}).
Subsequently, we again follow a procedure similar to that used in the
derivation of Eq.\ (\ref{Wsde1}) thus arriving at a general expression for
arbitrary momentum correlation functions,
\begin{widetext}
\begin{equation}
\Delta_{i_1\ldots i_n}^{p\ldots
p}=i^{n-1} \left[\left.\hat{\mathcal{O}}_{i_1\ldots i_{n-1}}^{p\ldots p}
\left( J^p_{i_n}+\frac{\delta\mathcal{S}_0}{\delta\dot{q}_{i_n}}\right)
\right\vert_{J^p=0}
\right]_{q\to \frac{\delta\mathcal{W}}{\delta
J^{q}}+\frac{\hbar}{i}\frac{\delta}{\delta J^{q}}}\label{mge}
\end{equation}
\end{widetext}
In contrast to the corresponding equation for the one-point function, Eq.\
(\ref{mge1}), (which did lead to the alternative set of DSEs) here the connected
Generating Functional in the standard Lagrange formalism appears. 
Note that the latter does not depend on $J_p$ anymore.
From the above functional equation, any other $m$-point Green's function with
$n$-external legs associated
to the $J^p$'s is obtained by taking $m\!-\!n$ derivatives with regard to
$J^q_{m}$ and setting them to zero.   In particular, the quantum average of
the momentum fields becomes a functional of the averaged field $\bar{q}$,
\begin{equation}
\bar{p}_{i}[\bar{q}]=\left.\frac{\delta\mathcal{S}_0}{\delta \dot{q}_i}
\right\vert_{q_m\to\bar{q}_i[J^q]+\frac{\hbar}{i}\Delta_{ij}^{qq}[J^q]
\frac{\delta}{\delta q_j}}\label{momenta}.
\end{equation}
Already at this point we want to point out that the averaged momentum field is 
generally not given by the usual definition of a canonical momentum field 
$\mathfrak{p}_i^{\mathrm{can}}$ on the level of the  Effective Action
\begin{equation}
\bar{p}_i[J=0]=\left< \frac{\delta S}{\delta \dot q} \right> \neq
\frac{\delta \Gamma }{\delta  \dot{\bar q}} \equiv 
\mathfrak{p}_i^{\mathrm{can}}
\, .
\end{equation}
This will be shown explicitly in subsection \ref{sub:proper-hamilton}.

In addition, the relation
between  the  $pp-$correlation functions and those that appear in
the standard formalism  can be expressed as
\begin{equation}
\Delta_{ij}^{pp}=\delta_{ij}+i\left.\left(\frac{\delta\mathcal{S}_0}{\delta
\dot{q}_{i}}\frac{\delta\mathcal{S}_0}{\delta\dot{q}_{j}}\right)
\right\vert_{q\to\bar{q}[J^q]+\frac{\hbar}{i}\Delta^{qq}[J^q]
\frac{\delta}{\delta
q}},\label{fomp}
\end{equation}
whereas the corresponding mixed $qp-$correlation function can be
 written as
\begin{equation}
\Delta_{ij}^{qp}[J^q]=\frac{\delta}{\delta
J^q_{i}}\left.\left(\frac{\delta\mathcal{S}_0}{\delta\dot{q}_{j}}
\right)\right\vert_{q_m\to\frac{\delta\mathcal{W}}{\delta
J^{q}}+\frac{\hbar}{i}\frac{\delta}{\delta J^{q}}}.\label{mix}
\end{equation}
By using the chain rule $\delta/\delta
J^q_{m}=\delta q_n/\delta J^q_{m}\delta/\delta q_{n}$, Eq.\ (\ref{mix})
can also be given in the form
\begin{equation}
\Delta_{ij}^{qp}[J^{q}]=\Delta_{il}^{qq}\frac{\delta}{\delta \bar{q}_l}
\left(\left.\frac{\delta\mathcal{S}_0}{\delta\dot{q}_j}
\right\vert_{q\to\bar{q}[J^q]+\frac{\hbar}{i}\Delta^{qq}[J^q]
\frac{\delta}{\delta q}}\right) =\Delta_{il}^{qq}
\frac{\delta {\bar p}_j}{\delta {\bar q}_l}. \label{fopp}
\end{equation}
The other mixed correlator $\Delta^{pq}$ can be immediately inferred from 
 the bosonic nature of
the fields, $\Delta_{ij}^{pq}= \Delta_{ji}^{qp}$.

Similarly, it is possible to determine the connected three point
correlation functions
\begin{eqnarray}
\Delta_{ijk}^{qpp}[J^{q}]&=&\Delta_{il}^{qq}\frac{\delta}{\delta
\bar{q}_l}\Delta_{jk}^{pp},\\
\Delta_{ijk}^{qqp}[J^{q}]&=&\Delta_{im}^{qq}\Delta_{jl}^{qq}\left
\{\frac{\delta^3\Gamma}{\delta\bar{q}_l\delta\bar{q}_m\delta\bar{q}_n}
\Delta_{nk}^{qp}+\frac{\delta^2\bar{p}_k}{\delta\bar{q}_m\delta\bar{q}_l}
\right\},\nonumber \\ \label{fopp1}\\
\Delta_{ijk}^{ppp}[J^{q}]&=&i\delta_{ij}\bar{p}_k+i\delta_{ik}\bar{p}_j
+i\delta_{jk}\bar{p}_i+\nonumber\\&+&i^2\left.\left(\frac{\delta\mathcal{S}_0}
{\delta
\dot{q}_{i}}\frac{\delta\mathcal{S}_0}{\delta\dot{q}_{j}}
\frac{\delta\mathcal{S}_0}{\delta\dot{q}_{k}}\right)
\right\vert_{q\to\bar{q}[J^q]+\frac{\hbar}{i}\Delta^{qq}[J^q]
\frac{\delta}{\delta q}}
\end{eqnarray}
where in the last expression we have used the previous results.

Employing  Eqs.\ (\ref{fopp}) and Eq.\ (\ref{fopp1})
the first and second derivatives of $\bar{p}$ with respect to
$\bar{q}$ can be expressed as functionals of the remaining elements.
In this way the vacuum expectation value of $p$
can be expanded. In fact,  up to second order in the classical field it
reads
\begin{eqnarray}
\bar{p}_m=&-&\frac{\delta^2\Gamma}{\delta\bar{q}_j\delta\bar{q}_l}
\Delta_{lm}^{qp}\bar{q}_j \nonumber\\
&-&\frac{1}{2!}\left[\frac{\delta^3\Gamma}
{\delta\bar{q}_i\delta\bar{q}_j\delta\bar{q}_l}\Delta_{lm}^{qp}-
\frac{\delta^2\Gamma}{\delta\bar{q}_i\delta\bar{q}_n}
\frac{\delta^2\Gamma}{\delta\bar{q}_j\delta\bar{q}_l}
\Delta_{lnm}^{qqp}\right]\bar{q}_i\bar{q}_j
\nonumber\\
&+& \ldots
\, .\label{mepn}
\end{eqnarray}
Since $\Delta_{ij}^{qp}$ can be computed using the procedure detailed above,
the expression (\ref{mepn}) determines $\bar{p}$ as a function of 
the second-order dressed correlation functions.

\subsection{Inverting the matrix propagator $\mathbb{D}$ }

The determination of the proper functions within the canonical formalism as a
functional of those appearing in the standard Lagrange framework is rather
cumbersome. This task involves connected tensors
of rank larger than two and depends on  the possibility to invert the propagator
$\mathbb{D}$. Once the individual elements of this propagator are computed, the
proper two-point function is completely determined in terms of the elements of
the Lagrange formalism.  In case the external sources associated to
the $p-$fields vanish, the inverse of $-\Delta^{qq}$ is  the proper Green's
function that arises in the standard path integral representation.

By direct inversion of the $2 \times 2$ block matrix one obtains
\begin{equation}
\mathbb{G}=\left(\begin{array}{cccc}
\Gamma_{ij}^{pp} &  \Gamma_{il}^{pp}\Delta_{lm}^{pq}\Gamma_{mj}^{qq}
\\   &   \\  \Gamma_{il}^{qq}\Delta_{lm}^{qp}\Gamma_{mj}^{pp}&
\left.\Gamma_{ij}^{qq}\right.^{H}\end{array}\right)\label{generalg}
\end{equation} where
\begin{eqnarray}
\Gamma_{ij}^{pp}=-\left(\Delta^{pp}+\Delta^{pq}\Gamma^{qq}
\Delta^{qp}\right)_{ij}^{-1}\label{ppifg}
\end{eqnarray}
and
\begin{equation}
\left.\Gamma_{ij}^{qq}\right.^{H}=\Gamma_{ij}^{qq}+\Gamma_{il}^{qq}
\Delta_{lk}^{qp}\Gamma_{km}^{pp}\Delta_{mn}^{pq}\Gamma_{nj}^{qq}.\label{shrepr}
\end{equation}

According to Eq.\ (\ref{generalg}) there are several equivalent
representations of the latter expression, {\it e.g.}
\begin{equation}
\left.\Gamma_{ij}^{qq}\right.^{H}=\left( \delta_{im} +\Gamma_{il}^{qp}
\Delta_{lm}^{pq}\right)\Gamma_{mj}^{qq}.\label{cucucu}
\end{equation}
However, in what follows we will consider the most simple form given by
\begin{equation}
\left.\Gamma_{ij}^{qq}\right.^{H}=\Gamma_{ij}^{qq}+\Gamma_{il}^{qp}
\left( \Gamma_{lm}^{pp} \right)^{-1} \Gamma_{mj}^{pq} ,\label{shreprqq}
\end{equation}
where we have introduced a unity in the form
$\mathbb{I}=\left( \Gamma^{pp} \right)^{-1} \Gamma^{pp}$ in Eq.\ (\ref{shrepr}),
and furthermore the expressions for the off-diagonal elements of  Eq.\
(\ref{generalg})  have been  used.

The method to obtain $\mathbb{G}$ can be generalized to any proper $m-$point
function. For instance, let us suppose that we want to compute the proper
three-point function. We denote it as $\mathbb{G}_{3}$ and the correspondingly
connected version as $\mathbb{D}^{(3)}.$  By considering the action of the
symbolic functional derivative $\delta/\delta J$ on Eq. (\ref{unidad}) we get
the equation
$\mathbb{D}^{(3)}\mathbb{G}+\mathbb{D}\mathbb{D}\mathbb{G}_{3}=0.$
By inversion of $\mathbb{D}$ in the second term  we obtain the desired form for
the proper 3-point Green's function $
\mathbb{G}_{3}=-\mathbb{G}\mathbb{G}\mathbb{D}^{(3)}\mathbb{G}.$  Proceeding in
an analogous way it is possible to express the proper four-point Green's
function $\mathbb{G}_{4}$, and so on. Thereby, the number of variables  within
the first order formalism can be reduced. From that point of view,  the
initially cumbersome problem becomes simpler.

\subsection{Connecting the Symmetry-Related identities \label{sec:srifs}}

Let us now return to the symmetry identity. To this end we write
Eq.\ (\ref{slavnov})  in the following form
\begin{eqnarray}
0&=&\int \mathcal{D}[q]\left\{J^q_{m}\delta q_m\left[\frac{\hbar}{i}
\frac{\delta}{\delta J^p},q\right]+J^p_{m}\delta p_m\left[\frac{\hbar}{i}
\frac{\delta}{\delta J^p},q\right]\right\}\nonumber\\
&\times&
\int \mathcal{D}[p]\exp\left[\frac{i}{\hbar}I[\phi,J]\right].
\label{importanteqq}
\end{eqnarray}
In order to simplify the following analysis let us consider theories where the
transformations are linear in the momentum fields.  Under this condition and
restricting to the class of Hamiltonians analyzed so far we obtain
\begin{eqnarray}
0&=&\int \mathcal{D}[q]\left\{J^q_{m}\delta q_m\left[J^p+
\frac{\delta \mathcal{S}_0}{\delta
\dot{q}},q\right] \right. \nonumber\\
&&+\left.J^p_{m}\delta p_m\left[J^p+\frac{\delta \mathcal{S}_0}{\delta
\dot{q}},q\right]\right\}\exp\left[\frac{i}{\hbar}\tilde{\mathcal{S}}[q,J]
\right].\label{importanteqqq}
\end{eqnarray}
In particular, if $J^p=0$ we find that the action $\mathcal{S}_0$ is invariant
under a symmetry transformation $\delta q_m$ which is a functional of $q$ only,
{\it i.e.} a symmetry transformation 
$\delta q_m\left[\frac{\delta\mathcal{S}_0}
{\delta\dot{q}},q\right] \to \delta q_m \left[ q \right]$.

The structure of Eq.\ (\ref{importanteqqq}) allows to write the
symmetry identities  in a similar form to Eq.\ (\ref{comparar1})
and Eq.\ (\ref{comparar2}),
\begin{eqnarray}
0&=&\int d^4x \left\{\frac{\delta \Gamma^H}{\delta \bar{p}_m}
\delta p_m\left[\frac{\delta \mathcal{S}}{\delta
\dot{q}}-\frac{\delta \Gamma^H}{\delta
\bar{p}},q\right]\right.\\
&&+\,\frac{\delta\Gamma^H}{\delta
\bar{q}_m} \left.\!\delta q_m\!\left[\frac{\delta \mathcal{S}_0}{\delta
\dot{q}}\!-\!\frac{\delta \Gamma^H}{\delta
\bar{p}},q\right]\right\}_{\phi\to
\bar{\phi}[J]+\frac{\hbar}{i}\Delta^{\phi\phi}[J]\frac{\delta}{\delta
\bar{\phi}}}. \nonumber
\end{eqnarray}
This concludes the formal discussion of the functional symmetry identities. 
A complete derivation, especially for constrained systems as Coulomb gauge 
Yang-Mills theories, is  presented  below in Sect.\ \ref{sec:coulomb}.

\section{Decomposition of proper Lagrange correlation functions 
\label{sec:lagrange-decomposition}}

In this section we will show how a general proper correlation function in the
Lagrange formalism can be decomposed into correlation functions in the Hamilton
approach.

\subsection{Relations between the bare elements}
So far we have presented the formalism for a general field theory.
In the present and forthcoming sections we will restrict ourselves to the most
important class of renormalizable field theories. In four dimensions the most
general renormalizable ``canonical action''  for a pure  bosonic theory can be
expressed as  a functional Taylor expansion,
\begin{eqnarray}
I_0[q,p]&=&I_{0ji}^{qp}p_iq_j+\frac{1}{2}I_{0ij}^{pp}p_ip_j+
\frac{1}{2}I_{0ijk}^{pqq}p_iq_jq_k \nonumber\\
&+&\frac{1}{2}I_{0ij}^{qq}q_iq_j
+\frac{1}{3!}I_{0ijk}^{qqq}q_iq_jq_k+
\frac{1}{4!}I_{0ijkl}^{qqqq}q_iq_jq_kq_l \, .\nonumber\\
\label{can-action}.
\end{eqnarray}
Here $I_{0ji}^{qp}=\partial_{\tau_i}\delta_{ij}$ is such  that
\begin{equation}
I_{0ji}^{qp}q_j=\dot{q}_i\label{dothg}.
\end{equation}
Clearly,  the  coefficients $I_{0i\ldots}^{\phi\ldots}$ are field
independent. They are given by the functional derivatives of $I_0$ evaluated
at $\phi=0,$ namely
\[
I_{0ij\ldots}^{\phi\phi}\equiv\left.\frac{\delta}{\delta\phi_i}
\frac{\delta}{\delta\phi_j}\frac{\delta}{\delta\phi_k}\ldots I_0
\right\vert_{\phi=0}.
\]
We remark that $I_{0ijk}^{pqq}$ as well $I_{0ijkl}^{qqqq}$ are  dimensionless
tensor couplings whereas $I_{0ij}^{qq}$ and $I_{0ijk}^{qqq}$ have mass
dimension $2$ and $1$, respectively, and do not involve time derivatives of
the fields. Nevertheless, depending on the assumed theory the latter two might
depend on $\nabla^2$ and $\nabla$. We are considering  bosonic theories, so
that all coefficients are symmetric. In particular, the first term in Eq.\
(\ref{acv}) can be written as $I_{0ji}^{pq}p_jq_i$ where
$I_{0ji}^{pq}=-\partial_{\tau_i}\delta_{ij}$ which leads to the functional
relation $\partial_{\tau_j}\delta_{ji}=-\partial_{\tau_i}\delta_{ij}.$

We can identify the  coefficients present in the Hamiltonian density Eq.\
(\ref{Hamiltoniand1}) as
\begin{equation}
\mathcal{A}_{ij}=-I_{0ij}^{pp}=\delta_{ij},\ \
\mathcal{B}_i=-\frac{1}{2}I_{0ijk}^{pqq}q_jq_k \label{ceff1}
\end{equation}
and
\begin{equation}
\mathcal{C}=-\frac{1}{2}I_{0ij}^{qq}q_iq_j-
\frac{1}{3!}I_{0ijk}^{qqq}q_iq_jq_k-\frac{1}{4!}I_{0ijkl}^{qqqq}q_iq_jq_kq_l.
\label{ceff2}
\end{equation}
Substituting Eqs.\ (\ref{ceff1}) and (\ref{ceff2}) in  Eq.\ (\ref{act})
and collecting the terms of the same order in $q$ we get that the action
$\mathcal{S}_0$ can be written as a polynomial functional,
\begin{equation}
\mathcal{S}_0=\frac{1}{2}\mathcal{S}_{0ij}q_iq_j+
\frac{1}{3!}\mathcal{S}_{0ijk}q_iq_jq_k+
\frac{1}{4!}\mathcal{S}_{0ijkl}q_iq_jq_kq_l.\label{a}
\end{equation}
In this context the following relations between the bare coefficients arise
\begin{equation}
\mathcal{S}_{0ij}\equiv \left. \frac{\delta^2\mathcal{S}_0}
{\delta q_i\delta q_j} \right|_{q=0}=
I_{0il}^{qp} I_{0lj}^{pq}+I_{0ij}^{qq} ,\label{bp}
\end{equation}
\begin{eqnarray}
\mathcal{S}_{0ijk} \equiv \left.\frac{\delta^3\mathcal{S}_0}
{\delta\bar{q}_i\delta\bar{q}_j\delta\bar{q}_k}\right|_{q=0}&=&
I_{0ikn}^{qqp} I_{0nj}^{pq}+\bar{q}_k\leftrightarrow
\bar{q}_j\mathrm{permut.}\nonumber\\&+&\bar{q}_i\leftrightarrow
\bar{q}_j\mathrm{permut.}+I_{0ijk}^{qqq},
\label{b3v}
\end{eqnarray}
\begin{eqnarray}
\mathcal{S}_{0ijkl} \equiv \left. \frac{\delta^4\mathcal{S}_0}
{\delta q_i\delta q_j\delta q_k\delta
q_l}\right|_{q=0}&=& I_{0ikn}^{qqp} I_{0njl}^{pqq}
+\bar{q}_i\leftrightarrow
\bar{q}_l\mathrm{perm.}\nonumber\\&+&\bar{q}_k\leftrightarrow
\bar{q}_l\mathrm{permut.}+I_{0ijkl}^{qqqq} . \nonumber \\ \label{b4v}
\end{eqnarray}
Substituting the above equations in Eq.\ (\ref{a}) and considering the
relation  given by Eq.\ (\ref{dothg})  allows to find additional relations.
In fact, taking the second, third and fourth functional derivatives of
$\mathcal{S}_0$ and evaluating at $q=0$, respectively, we obtain
\begin{equation}
\mathcal{S}_{0ij}^{q\dot q} \equiv \left. \frac{\delta^2\mathcal{S}_0}
{\delta q_i\delta
\dot{q}_j} \right|_{q=0}=I_{0ij}^{qp}=\partial_{\tau_j}\delta_{ji}
\label{lamarmotadormida}
\end{equation}
\begin{equation}
\mathcal{S}_{0ji}^{\dot q q} \equiv \left.
\frac{\delta^2\mathcal{S}_0} {\delta \dot{q}_j\delta q_i}
\right|_{q=0} = I_{0ij}^{pq} = -\partial_{\tau_j}\delta_{ji},
\label{lamarmotadormida21}
\end{equation}
\begin{equation}
\mathcal{S}_{0ijk}^{qq\dot q} \equiv \left. \frac{\delta^3 \mathcal{S}_0}
{\delta q_i\delta q_j\delta \dot{q}_k} \right|_{q=0} =I_{0ijk}^{qqp},
\label{lamarmota}
\end{equation}
\begin{equation}
\left.\frac{\deltabar^3 \mathcal{S}_0}{\delta q_i\delta q_j \delta
q_k}\right|_{q=0}=I_{0ijk}^{qqq}\ \ \textrm{and} \
\ \left.\frac{\deltabar^4 \mathcal{S}_0}
{\delta q_i\delta q_j \delta q_k\delta q_l}\right|_{q=0}=I_{0ijkl}^{qqqq}.
\end{equation}
In the last two relations we have introduced $\deltabar$  denoting  a partial
functional differentiation  acting just  on those terms in the
action  that do not involve the time derivative of the field.

To complete our analysis we  point out that from Eq.\ (\ref{cvbr})
the general form of the quantum canonical momentum fields in four-dimensional
renormalizable theories  is given by
\begin{eqnarray}
p_i=\mathcal{S}_{0ji}^{q\dot q} q_j+
\frac{1}{2} \mathcal{S}_{0kji}^{qq\dot q} q_jq_k.\label{gqcm}
\end{eqnarray}

As these polynomial representations allow to identify a priori the bare
elements,
the expansions given above prove to be very convenient to derive the
DSEs for a general bosonic theory in both formulations.

\subsection{Diagrammatic representation\label{graphical-rep}}

The relations between the correlation functions in both formulations will be
given in the following via explicit diagrammatic expressions. To enable this,
we will first introduce a graphical representation in terms of the
fundamental objects that characterize the theory in both formulations.

\begin{itemize}
\item[I)]{As before we consider fields that involve all the
irreducible representations in the theory. The fundamental fields are represented by
solid lines whereas the corresponding momentum fields are denoted by
zigzag lines.
\begin{center}
\includegraphics[width=0.3\textwidth]{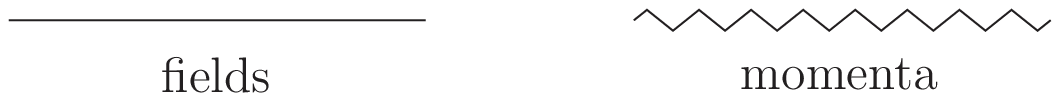}
\end{center}}

\item[II)]{Dressed propagators are denoted by thick, whereas bare propagators
and external lines by thin lines, respectively. Off-diagonal propagator
components are represented by a thick, half solid and half zigzag line.
\begin{center}
\includegraphics[width=0.3\textwidth]{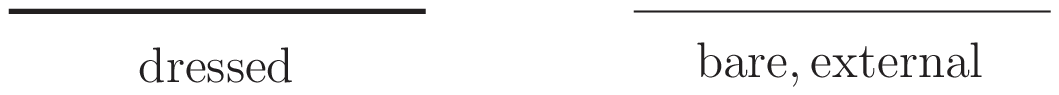}
\end{center}}
To keep the representation of the DSEs in the first order formalism
concise we also include the matrix propagator Eq.\ (\ref{cpro}) represented
by a double line.
\begin{center}
\includegraphics[width=0.3\textwidth]{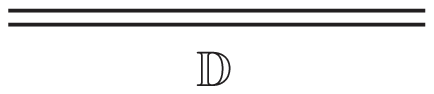}
\end{center}

\item[III)]{All proper correlation functions (including proper 2-point
functions) in the first and second order formalism are denoted by small and
large filled blobs,
\begin{center}
\includegraphics[width=0.3\textwidth]{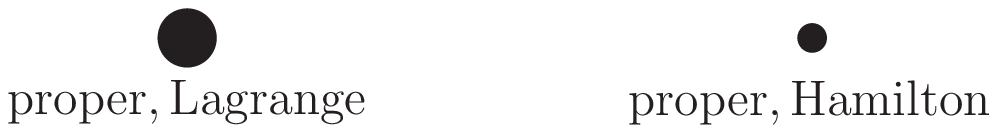}
\end{center}
whereas bare vertex functions are represented by open blobs.
\begin{center}
\includegraphics[width=0.3\textwidth]{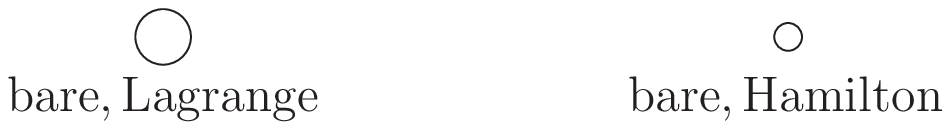}
\end{center}}

\item[IV)]{In our analysis it will become useful to introduce the inverse
proper momentum 2-point function
\begin{equation}
\mathcal{D}^{pp}_{ij}\equiv-\left(\frac{\delta^2\Gamma^H}
{\delta\bar{p}_i\delta\bar{p}_j}\right)^{-1}
\end{equation}
which will be represented by a thick dotted line. Since it has the form of an
alternative momentum propagator it can connect as an internal momentum line to
proper vertices. Similarly all connected correlators which are
one-particle-irreducible (1PI) in the
fields but merely connected via the "propagator" $\mathcal{D}^{pp}$ are
called {\em p-connected} and are denoted by a blob labeled by a P.
\begin{center}
\includegraphics[width=0.3\textwidth]{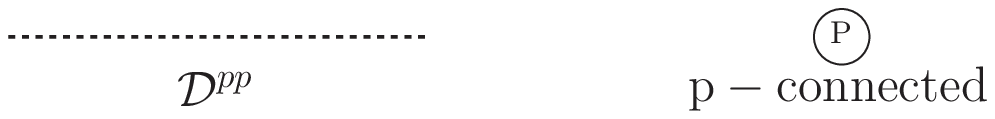}
\end{center}}
\end{itemize}

\subsection{Diagrammatic decomposition\label{sub:proper-decomposition}}

To express proper functions in the Lagrange formalism in terms of those of the
Hamilton formalism, we exploit the underlying equivalence between the first
and second order formalism.
Due to the equivalence of the Generating Functionals of connected Green's
functions at vanishing sources $J^p$, the Effective Actions in the two
formalisms are identical when $\bar p$ is a functional of $\bar q$ which in 
turn is
implicitly given by the stationarity in $\bar p$,
\begin{equation}
\Gamma[\bar{q}]\equiv\Gamma^H\left[\bar{p}(\bar{q}),\bar{q}\right]
\label{equivalenciaquantica}\
\ \mathrm{whenever}\ \ \frac{\delta \Gamma^H}{\delta\bar{p}}=-J^p=0.
\end{equation}
Here $\Gamma^H$ still depends explicitly on  $\dot{\bar{q}}$ and $\bar{p}$,
where the quantum average of the momentum fields is in general a complicated
functional of the fundamental field.
Any proper $n$-point  function in the standard formalism can be
determined by  taking $n$ derivatives with respect to the fields
in Eq.\ (\ref{equivalenciaquantica}) and evaluated at the vacuum expectation
value. In particular applying the chain rule and Eq.\
(\ref{equivalenciaquantica}) the first derivative reads
\begin{equation}
\frac{\delta \Gamma}{\delta \bar{q}_i}=
\frac{\delta \Gamma^H}{\delta \bar{q}_i}+
\frac{\delta \Gamma^H}{\delta \bar{p}_j}
\frac{\delta \bar{p}_j}{\delta \bar{q}_i}=
\frac{\delta \Gamma^H}{\delta \bar{q}_i} . \label{opstart}
\end{equation}
In the above defined graphical representation this equation reads
\begin{center}
\includegraphics[width=0.2\textwidth]{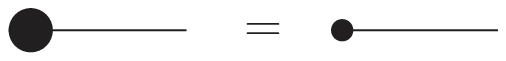} .
\end{center}
As usual in the standard formalism, a differentiation with respect to a
field is equivalent to attaching an external leg in the graphical
representation
({\it cf., e.g.,} ref.\  \cite{Alkofer:2008jy}).
Next we consider the second derivative of the action given by
\begin{equation}
\frac{\delta^2\Gamma}{\delta\bar{q}_i \delta\bar{q}_j}=
\frac{\delta\bar{p}_n}{\delta\bar{q}_i}
\frac{\delta^2\Gamma^H}{\delta\bar{p}_n
\delta\bar{q}_j}+\frac{\delta^2\Gamma^H}{\delta\bar{q}_i
\delta\bar{q}_j} \, .\label{fundamenta2}
\end{equation}
The field derivative of the momentum field can alternatively to the previous
result (\ref{mepn}) also be obtained from a field derivative of the
constraint equation in Eq.\ (\ref{equivalenciaquantica}),
\begin{eqnarray}
\frac{\delta^2\Gamma^H}{\delta\bar{q}_i\delta\bar{p}_j}=
-\frac{\delta\bar{p}_n}{\delta\bar{q}_i}
\frac{\delta^2\Gamma^H}{\delta\bar{p}_n\delta\bar{p}_j}\;
\Rightarrow \; 
\nonumber \\
\frac{\delta\bar{p}_n}{\delta\bar{q}_i} =
- \frac{\delta^2\Gamma^H}{\delta\bar{q}_i\delta\bar{p}_j}
\left( \frac{\delta^2\Gamma^H}{\delta\bar{p}_n\delta\bar{p}_j} \right)^{-1} .
\label{complementary1}
\end{eqnarray}
Inserting this in Eq.\ (\ref{fundamenta2}) it is expressed entirely in terms
of proper first order Green's functions and takes the symmetric form
\begin{equation}
\frac{\delta^2\Gamma}{\delta\bar{q}_i \delta\bar{q}_j}=\frac{\delta^2\Gamma^H}
{\delta\bar{q}_i
\delta\bar{q}_j} - \frac{\delta^2\Gamma^H}{\delta\bar{q}_i\delta\bar{p}_j}
\left( \frac{\delta^2\Gamma^H}{\delta\bar{p}_n\delta\bar{p}_j} \right)^{-1}
\frac{\delta^2\Gamma^H}{\delta\bar{p}_n
\delta\bar{q}_j} 
\end{equation}
in accordance with Eq. (\ref{shreprqq}). The above equation yields the graphical 
representation of the decomposition of the proper two-point Green's function
\begin{center}
\includegraphics[width=0.45 \textwidth]{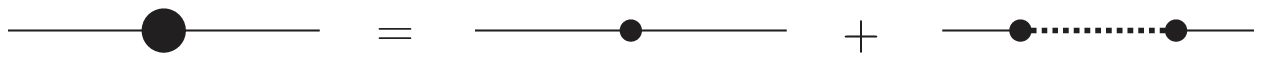} .
\end{center}
Interestingly, in terms of the first order correlation functions there is
in addition to the proper part also a connected contribution due to the fact
that the fundamental field mixes with the corresponding momentum field. This is
evident from the off-diagonal elements Eq.\ (\ref{mix}) in the propagator
matrix in the first order formalism. Due to the mixing the arising propagator
in Eq.\ (\ref{complementary1}) is not the elementary $p$-propagator but the
propagator for a collective mode described by the inverse of the proper
two-point momentum correlation function $\mathcal{D}^{pp}$
which is represented by the dotted line and related to the actual momentum
propagator via Eq.\ (\ref{ppifg}).

The result for the fundamental field derivative of a proper correlation
function in the  Hamilton formalism yields the replacement rule
\begin{center}
\includegraphics[width=0.35 \textwidth]{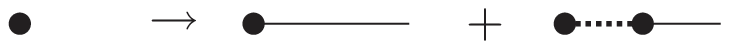} .
\end{center}
In the next step a field derivative can also act on the  propagator
$\mathcal{D}^{pp}$. Its derivative is obtained from the derivative of the
inverse of an operator as
\begin{eqnarray}
\frac{\delta \mathcal{D}_{ij}^{pp}}{\delta \bar{q}_k} &=&
-  \frac{\delta}{\delta \bar{q}_k}
\left( \frac{\delta^2\Gamma^H}{\delta\bar{p}_i\delta\bar{p}_j} \right)^{-1} 
\label{propder} \\
&=& \left( \frac{\delta^2\Gamma^H}{\delta\bar{p}_i\delta\bar{p}_m} \right)^{-1}
\!\!\! \frac{\delta^3\Gamma^H}{\delta\bar{p}_m\delta\bar{p}_n \delta \bar{q}_k}
\! \left( \frac{\delta^2\Gamma^H}{\delta\bar{p}_n\delta\bar{p}_j} \right)^{-1}
\nonumber \\
&=& \mathcal{D}^{pp}_{im} \Gamma^{ppq}_{mnk} \mathcal{D}^{pp}_{nj}
\nonumber
\end{eqnarray}
which yields the graphical replacement rule
\begin{center}
\includegraphics[width=0.35 \textwidth]{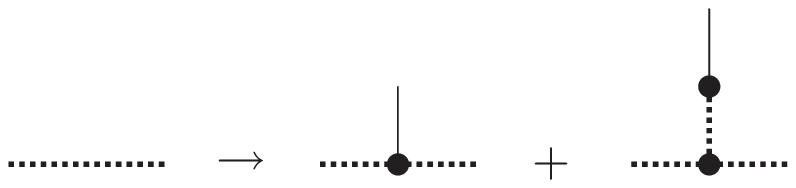} .
\end{center}
Applying these two replacement rules in all possible ways on the right hand
side of the above equation for the two-point vertex provides immediately the
corresponding decomposition of the proper 3-point vertex
\begin{center}
\includegraphics[width=0.45 \textwidth]{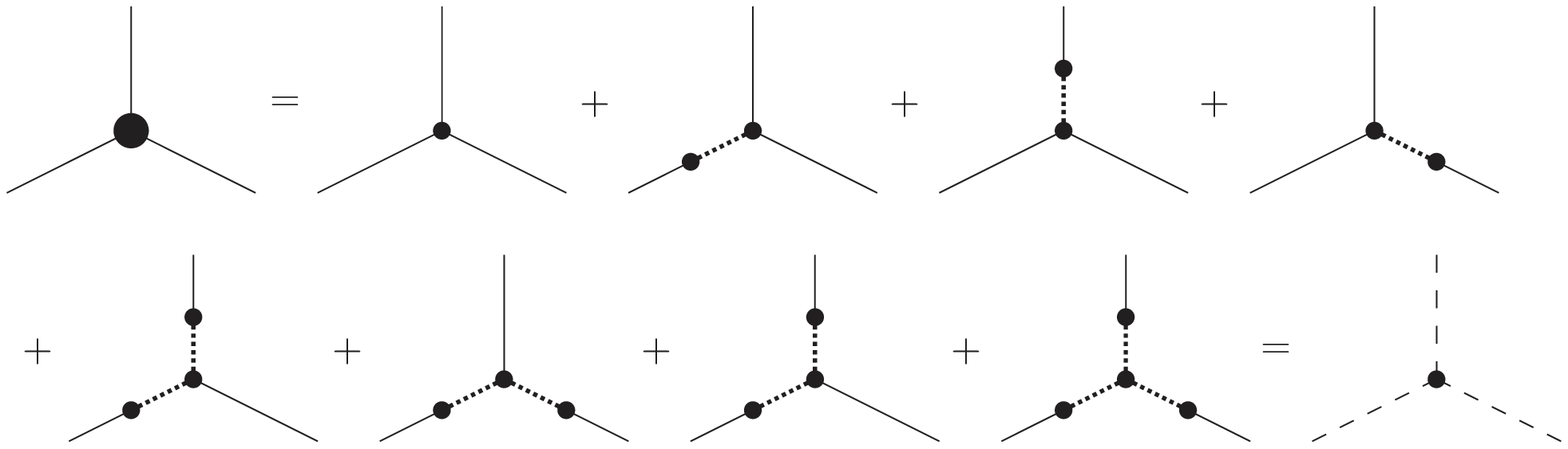} .
\end{center}
This yields directly a symmetric result, whereas the computation without
the replacement Eq.\ (\ref{complementary1}) in each step would produce an
asymmetric result involving higher derivatives of $\bar{p}$.

\begin{figure}
\begin{center}
\includegraphics[width=0.2\textwidth]{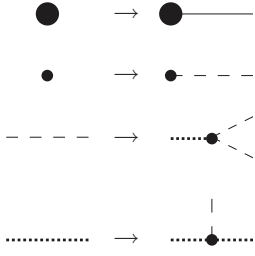}
\end{center}
\vspace*{-0.5cm}
\caption{The replacement rules that create the general decomposition of a
proper correlation function in the Lagrange formalism in terms of Hamilton
correlation functions.\label{fig:rules}}
\end{figure}

To derive the decomposition of higher order Lagrange correlation functions
it is useful to note that both of the above replacement rules involve the
external legs in the form of the composite expression
\begin{center}
\includegraphics[width=0.35 \textwidth]{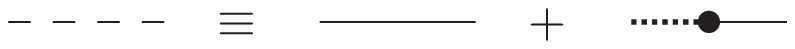}
\end{center}
denoted by a dashed line. By introducing the composite external leg into the
graphical representation, the decomposition of the 3-point function is given
by a single graph with three of these new external legs.  Via the previous
rules it is easy to obtain a corresponding replacement rule for the composite
external leg.  Thereby the extension of a general $n$-point function by an
additional leg can be obtained from the simplified set of rules in
Fig.~\ref{fig:rules}.
Starting from the 3-point function these rules allow to derive the
decomposition of arbitrary proper $n$-point functions in the Lagrange
formalism. In the appendix we show that the graphs generated by these
replacement rules have a very simple structure that can be summarized by the
following general statement:
{\quotation \em
A proper $n$-point function in the Lagrange formalism can be decomposed into
the sum of all $p$-connected $n$-point functions with composite external legs
in the Hamilton framework.}

\medskip

\begin{figure*}
\begin{center}
\includegraphics[width=0.7 \textwidth]{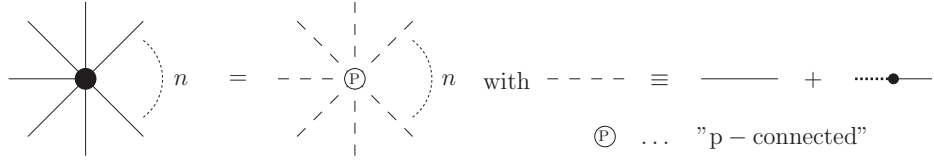}
\end{center}
\caption{The general result for the decomposition of a proper correlation 
function in the Lagrange framework in terms of correlators in the Hamilton 
formulation.\label{fig:Lag-dec}}
\end{figure*}

\noindent
This is shown in graphical form in Fig. \ref{fig:Lag-dec}.

The graphical representation manifestly shows that the propagation, decay and
dispersion of particles are more complicated when analyzed in the Hamilton
framework. At the tree level the arising propagators $\mathcal{D}^{pp}$ are
constant and correspondingly the dotted internal lines between the proper
vertices represent merely contact terms. Similarly the external leg corrections
involve no additional poles at tree level and the explicit energy dependence
arising from the mixed correlator in Eq.\ (\ref{complementary1}) just cancels
the one from the momentum field derivative of the initial vertex. In the fully
dressed action, however, there could arise additional poles due to the actual
propagation of the momentum fields via higher order kinetic terms, as well
as an explicit energy dependence of the terms that cancels only after summing
over all contributions.

Via graphical rules, one can therefore  generate in a systematical way the 
connection between any proper Green's function in the standard second-order
formalism to the appropriate correlation functions in the first-order formalism.
We point out already here that the inverse relations for Green's functions in
the latter context are more complicated and can involve loops  related to
connected correlation functions with both mixed and pure momentum fields. The
method used in the last  section allows us to represent them in terms of those
appearing in the Lagrange formalism.

\subsection{Inclusion of Grassmannian fields}

As already mentioned we will consider for completeness also 
quantum field theories involving Grassmannian fields
$c_i$ and $c_j^\dagger$ fulfilling the anti-commutation rules
\begin{eqnarray}
\left\{c_i,c_j^\dagger\right\}&=&\delta_{ij} \, ,\\
\left\{c_i^\dagger,c_j^\dagger\right\}&=&\left\{c_i,c_j\right\}=0.
\end{eqnarray}
The Grassmannian  action for a renormalizable theory in four dimensions has the
general form
\begin{equation}
I_0=i\kappa_{ij}c_i^\dagger\dot{c}_{j}+i\lambda_{ijk}q_ic^{\dagger}_jc_k
+i\alpha_{ij}c_i^\dagger c_{j}.\label{fp2}
\end{equation}
The quantum canonical momentum fields
associated to the $c_i$ are given by
$p_i=\frac{\delta
I_0}{\delta_\mathrm{R}\dot{c}_i}=i\kappa_{ij}c^\dagger_j$, where the
suffix \emph{R} denotes differentiation from the right. It is clear that
the usual path integral representation for Grassmannian field is already of 
first-order
form since the momentum fields are treated as independent variables.
{\em  Purely fermionic
correlation functions are therefore trivially identical in the two
formulations.}

Let us now study mixed correlation functions involving bosonic fields
represented by the $q_i$ in Eq.\ (\ref{fp2}) when the bosonic path integral
is written in the canonical form as well. Generally the tensors
$\kappa_{ij},$ $\alpha_{ij}$ and
$\lambda_{ijk}$ are real and field independent.
The kinetic tensor reads in particular
\[
\kappa_{ij}=\left\{\begin{array}{cccc}
  \delta_{ij} &  \mathrm{with}\ \ {c, c^\dagger} \in \mathrm{Fermions} \\
   &   \\
  \delta_{ij}\sigma\partial_0 & \mathrm{with}\ \ {c, c^\dagger} \in \mathrm{Fadeev-Popov\;\; Ghosts}
\end{array}\right.
\] where $\sigma\in\mathbb{R}.$
Since it is a bosonic variable, $\bar{p}$ is a functional of fermionic
bilinears $\bar{c_i}^\dagger \Gamma_{ij} \bar{c_j}$ and the field
$\bar{q}$.  As a consequence  the second term in Eq.\ (\ref{fundamenta2})
vanishes identically, which confirms  the equality of the
propagators in both formulations
\begin{equation}
\frac{\delta^2\Gamma}{\delta \bar{c}_i\delta \bar{c}_j^\dagger}=
\frac{\delta^2\Gamma^H}{\delta \bar{c}_i\delta \bar{c}_j^\dagger}.
\end{equation}
Analogous to the bosonic case the three-point vertex can be decomposed as
\begin{equation}
\frac{\delta^3\Gamma}{\delta\bar{q}_i\delta \bar{c}_j\delta \bar{c}_k^\dagger}=
\frac{\delta\bar{p}_n}{\delta  \bar{q}_i}
\frac{\delta^3\Gamma^H}
{\delta \bar{p}_n\delta \bar{c}_j\delta \bar{c}_k^\dagger}+
\frac{\delta^3\Gamma^H}
{\delta \bar{q}_i\delta \bar{c}_j\delta \bar{c}_k^\dagger}.
\end{equation}
Although the propagators are the same in both formulations the vertices can
be different. The above chain rule again results in the appearance of composite
legs for all bosons in the graphical representation. The decomposition of a
general $n$-point correlation function is then again given by all p-connected
graphs
where only the bosonic propagators $\mathcal{D}^{pp}$ are involved and only
bosonic external legs are composite.

\section{Decomposition of connected Hamilton correlation functions
\label{sec.con-ham}}

So far we gave in subsection \ref{gen-Ham-con} general expressions for
correlation functions in the Hamilton formalism. The goal in this section is
to evaluate the general decomposition of the two-point functions in the
Hamilton framework in case of a generic four-dimensional renormalizable
quantum field theory in terms of Lagrange correlation functions. To do this,
we start by computing the elements of $\Delta_{ij}^{qp}.$

\subsection{The mixed connected 2-point function}
\begin{figure}[h]
\includegraphics[width=0.47\textwidth ]{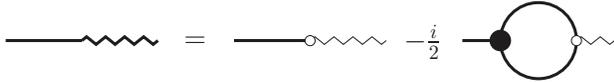}
\caption{The decomposition of the mixed Hamilton propagator in terms of 
Lagrange correlation functions.\label{fig:mixed-dec}}
\end{figure}

By considering Eq. (\ref{fopp}) and Eq. (\ref{gqcm}) the mixed connected
two-point function in the first order formalism  can be written as
(from now on we skip the explicit factor $\hbar$)
\begin{equation}
\Delta_{ij}^{qp}=\Delta_{il}^{qq}\left(\mathcal{S}_{0lj}^{q
\dot{q}}-\frac{i}{2}\mathcal{S}_{0ukj}^{qq\dot{q}}
\frac{\delta}{\delta\bar{q}_l}\Delta_{uk}^{qq}\right).
\end{equation}
Using partial differentiation for the identity
$\Delta_{il}^{qq}\Gamma_{lj}^{qq}=-\delta_{ij}$ we obtain
\begin{equation}
\frac{\delta}{\delta\bar{q}_l}\Delta_{uk}^{qq}
=\Delta_{um}^{qq}\Gamma_{mln}^{qqq}\Delta_{nk}^{qq} ,\label{const}
\end{equation}
and therefore in configuration space 
\begin{equation}
\Delta_{ij}^{qp}=\Delta_{il}^{qq}\left(I_{0lj}^{
qp}-\frac{i}{2}I_{0ukj}^{qqp}\Delta_{um}^{qq}\Gamma_{mln}^{qqq}\Delta_{nk}^{qq}
\right),\label{qppp}
\end{equation}
where  Eq.\ (\ref{lamarmotadormida}) and Eq.\ (\ref{lamarmota}) have been used.
According to the diagrammatic representation the above function has the
decomposition given in Fig.~\ref{fig:mixed-dec}.

Via the bosonic symmetry of the propagator it is easy to see that
\begin{equation}
\Delta_{ij}^{pq}=\left(I_{0li}^{
qp}-\frac{i}{2}I_{0uki}^{qqp}\Delta_{um}^{qq}\Gamma_{mnl}^{qqq}
\Delta_{nk}^{qq}\right)\Delta_{lj}^{qq}.\label{pppq}
\end{equation}
Clearly, Eq.\ (\ref{qppp}) and Eq.\ (\ref{pppq}) fulfill the condition
$\Delta_{ij}^{qp} = \Delta_{ji}^{pq}$.

Our convention for  the Fourier transform of a general two-point function
(connected or proper) obeying translational invariance is
\begin{equation}
\Delta_{ij}^{\Phi\Phi}\equiv\Delta_{ij}^{\Phi\Phi}(x_i-x_j)=
\int \dbar k \Delta_{ij}^{\Phi\Phi}(k)e^{-ik(x_i-x_j)},
\end{equation}
with $k(x_i-x_j)=k_0(x_{0i}-x_{0j})-\vec{k}\cdot(\vec{x}_{i}-\vec{x}_{j})$ and
$\dbar k\equiv  d^4k/(2\pi)^4$.
As a consequence of this  convention and the equivalence
$\Delta_{ij}^{qp} = \Delta_{ji}^{pq}$ we obtain the relation
\begin{equation}
\Delta_{ij}^{qp}(k)=\Delta_{ji}^{pq}(-k) \label{cosabuena}
\end{equation}
where $i$ and $j$ represent the remaining internal indices.
This yields the corresponding equation in momentum space
\begin{eqnarray}
\Delta_{ij}^{qp}(k)&=&\Delta_{il}^{qq}(k)\left(
ik_0\delta_{lj}\!-\!\frac{i}{2}
\!\int\!\dbar\omega I_{0ukj}^{qqp}(k-\!\omega\!,\omega,-k)
\right.\nonumber\\&\times&\left.\Delta_{um}^{qq}(\omega-k)
\Gamma_{mln}^{qqq}(\omega-k,k,-\omega)\Delta_{nk}^{qq}(\omega)\right).
\nonumber\\
\label{sadad}
\end{eqnarray}
In the derivation of the latter equation we have taken into account
the Fourier transformation of the proper 3-point function
\begin{eqnarray}
\Gamma_{\alpha\beta\gamma}^{\Phi\Phi\Phi}&=&\int\dbar k_\alpha\ \
\dbar k_\beta\ \ \dbar k_\gamma\left(2\pi\right)^{4}
\delta^{(4)}(k_\alpha+k_\beta+k_\gamma)\nonumber\\&
\times&\Gamma_{\alpha\beta\gamma}^{\Phi\Phi\Phi}(k_\alpha,k_\beta,k_\gamma)
e^{-ik_\alpha x_\alpha-ik_\beta x_\beta-ik_\gamma x_\gamma}
\end{eqnarray}
where the $\delta-$function expresses the momentum conservation.

\subsection{The momentum propagator \label{sub:momprop}}

The pure $pp-$correlator can be computed using Eqs.\ (\ref{fomp}) and
(\ref{gqcm}). Neglecting those terms that will eventually vanish when the
sources are set to zero, it reads
\begin{eqnarray}
\Delta_{ij}^{pp}&=&\delta_{ij}+\mathcal{S}_{0li}^{q\dot{q}}
\mathcal{S}_{0mj}^{q\dot{q}}\Delta_{lm}^{qq}-
\frac{i}{2}\mathcal{S}_{0li}^{q\dot{q}}\mathcal{S}_{0nmj}^{qq\dot{q}}
\Delta_{lu}^{qq}\frac{\delta}{\delta\bar{q}_u}\Delta_{nm}^{qq}\nonumber\\
&-&\frac{i}{2}\mathcal{S}_{0mj}^{q\dot{q}}\mathcal{S}_{0kli}^{qq\dot{q}}
\Delta_{ku}^{qq}\frac{\delta}{\delta\bar{q}_u}\Delta_{lm}^{qq}-\frac{i}{2}
\mathcal{S}_{0kli}^{qq\dot{q}}\mathcal{S}_{0nmj}^{qq\dot{q}}\Delta_{km}^{qq}
\Delta_{ln}^{qq}\nonumber\\&-&\frac{1}{4}\mathcal{S}_{0kli}^{qq\dot{q}}
\mathcal{S}_{0nmj}^{qq\dot{q}}\Delta_{ku}^{qq}\frac{\delta}{\delta\bar{q}_u}
\left[\Delta_{lx}^{qq}\frac{\delta}{\delta\bar{q}_x}\Delta_{nm}^{qq}\right].
\end{eqnarray}
By iterated application of Eq.\ (\ref{const}) and considering Eqs.\
(\ref{lamarmotadormida}) and (\ref{lamarmota}) this expression yields
\begin{widetext}
\begin{eqnarray}
\Delta_{ij}^{pp}&=&\delta_{ij}-\frac{i}{2}I_{0ikl}^{pqq} \Delta_{km}^{qq}
\Delta_{ln}^{qq}I_{0nmj}^{qqp}-\frac{1}{4}I_{0ikl}^{pqq}\Delta_{ku}^{qq}
\Delta_{lx}^{qq}\Delta_{ny}^{qq}\Gamma_{yxuz}^{qqqq}
\Delta_{zm}^{qq}I_{0nmj}^{qqp}-\frac{1}{2}I_{0ikl}^{pqq}\Delta_{ku}^{qq}
\Delta_{lx}^{qq}\Delta_{np}^{qq}\Gamma_{puq}^{qqq}\Delta_{qy}^{qq}
\Gamma_{yxz}^{qqq}\Delta_{zm}^{qq}I_{0nmj}^{qqp}\nonumber\\
&&+I_{0il}^{pq} \Delta_{lm}^{qq} I_{0mj}^{qp} - \frac{i}{2} I_{0il}^{pq}
\Delta_{lm}^{qq}\Gamma_{xmy}^{qqq} \Delta_{xn}^{qq} \Delta_{yu}^{qq}
I_{0nuj}^{
qqp} - \frac{i}{2} I_{0ikl}^{
pqq} \Delta_{ku}^{qq}\Delta_{lx}^{qq}\Gamma_{xuy}^{qqq} \Delta_{ym}^{qq}
I_{0mj}^{qp} \nonumber \\
&&-\frac{1}{4} I_{0ikl}^{
pqq} \Delta_{ku}^{qq}\Delta_{lx}^{qq}\Gamma_{xuy}^{qqq} \Delta_{ym}^{qq}
\Delta_{nf}^{qq}\Gamma_{fmz}^{qqq}\Delta_{zh}^{qq}  I_{0nhk}^{
qqp} \, . \label{ppexpf}
\end{eqnarray}
Here we have already suppressed disconnected expressions involving tadpoles
that cancel since Eq.\ (\ref{mge11}) gives
\begin{equation}
\left.\frac{\delta\Gamma^H}{\delta\bar{p}_i}\right|_{\bar{q},\bar{p}=0}
\sim I_{0mni} ^{qqp}\Delta_{mn}^{qq}=0 \, .
\end{equation}
The graphical representation of the decomposition of the momentum propagator
in the Hamilton framework is shown in Fig. \ref{fig:mom-dec}.
\begin{figure*}
\begin{center}
\includegraphics[width=.9\textwidth]{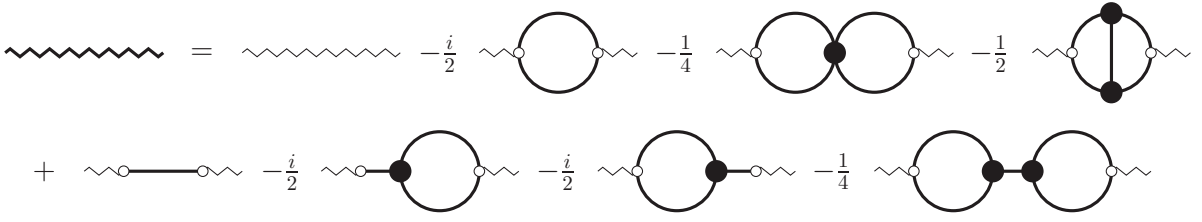}
\end{center}
\caption{The decomposition of the momentum propagator in terms of Lagrange 
correlation functions.\label{fig:mom-dec}}
\end{figure*}
Note that the loop graphs in the first line represent precisely the vacuum 
graphs of the $n$-particle irreducible ($n$PI) actions in the Lagrange 
formalism of order one and two with attached external $p$-legs. In addition to 
these proper contributions this
result involves again connected graphs in the second line. These graphs
correspond to the second and third line in Eq.\ (\ref{ppexpf})   
which can
alternatively be expressed by the known mixed correlation functions
derived in the last subsection as
$-\Delta_{im}^{pq}\Gamma_{ml}^{qq}\Delta_{lj}^{qp}$ and
correspondingly represented as a single graph.

After Fourier transformation of Eq.\ (\ref{ppexpf})  
we find the representation
of $\Delta^{pp}$ in momentum space
\begin{eqnarray}
\Delta_{ij}^{pp}(k)&=&\delta_{ij}-\Delta_{im}^{pq}(k)\Gamma_{ml}^{qq}(k)
\Delta_{lj}^{qp}(k)-\frac{i}{2}\int
\dbar\omega
I_{0ikl}^{pqq}(k,\omega-k,-\omega)I_{0nmj}^{qqp}(\omega,k-\omega,-k)
\Delta_{km}^{qq}(\omega-k)\Delta_{ln}^{qq}(\omega)\nonumber\\
&-&\frac{1}{4}\int\dbar\omega\dbar
\mu
I_{0ikl}^{pqq}(k,\mu-k,-\mu)I_{0nmj}^{qqp}(-\omega,\omega+k,-k)
\Delta_{ku}^{qq}(\mu-k)\Delta_{lx}^{qq}(-\mu)\Delta_{ny}^{qq}(-\omega)
\Gamma_{yxuz}^{qqqq}(\omega,\mu,k-\mu,-\omega-k)\nonumber\\
&\times&\Delta_{zm}^{qq}(-\omega-k)-\frac{1}{2}\int\dbar\omega\dbar
\mu
I_{0ikl}^{pqq}(k,\omega-k,-\omega)I_{0nmj}^{qqp}(\mu+k,-\mu,-k)
\Delta_{ku}^{qq}(\omega-k)\Gamma_{puq}^{qqq}(-k-\mu,k-\omega,\omega+\mu)
\nonumber\\
&\times&\Delta_{lx}^{qq}(-\omega)\Delta_{np}^{qq}(\mu+k)
\Delta_{qy}^{qq}(\omega+\mu)\Gamma_{yxz}^{qqq}(-\omega-\mu,\omega,\mu)
\Delta_{zm}^{qq}(\mu).
\label{ppexpf1}
\end{eqnarray}
\end{widetext}

The summation over all field components indicated by the subindices
in Eq.\ (\ref{sadad}) (see also Eq. (\ref{ppexpf}))  
leads to multiple
possibilities. Yet many of these might not be allowed due to the
symmetries of the theory. The
decomposition of higher order functions involving one or two
external momentum fields is simply obtained by further derivatives
of these generating equations. Via the chain rule a derivative
w.r.t.\ to the source can be transformed into a derivative with
respect to the averaged fields yielding in addition external
propagators. Correspondingly the graphical decomposition of such
correlation functions can be obtained recursively via the graphical
replacement rules in the Lagrange framework given in
\cite{Alkofer:2008jy}. On the other hand, from Eq.\ (\ref{mge}) it
is clear that higher order correlation functions involving $n$
momentum fields involve graphs of loop order $n$. Therefore it is not
possible to give a close result for all $n$-point functions.

To conclude this section we remark that the multiplication of $\Gamma^{qq}$
on the left hand side  amputates the Green's function $\Delta^{qq}.$  According
to Eq.\ (\ref{mepn}), this operation allows to derive
\begin{equation}
\frac{\delta \bar{p}_i}{\delta \bar{q}_j}=I_{0ji}^{
qp}-\frac{i}{2}I_{0uki}^{qqp}\Delta_{um}^{qq}\Gamma_{mjn}^{qqq}\Delta_{nk}^{qq}.
\end{equation}
Considering this expression and the diagrammatic rules
in the standard path integral representation we get
\begin{eqnarray}
\frac{\delta
\bar{p}_i}{\delta\bar{q}_j\delta\bar{q}_k}&=&I_{0jki}^{qqp}-
\frac{i}{2}I_{0mui}^{qqp}\Delta_{mn}^{qq}\Gamma_{nkx}^{qqq}\Delta_{xy}^{qq}
\Gamma_{yjt}^{qqq}\Delta_{tu}^{qq}+j\leftrightarrow k\nonumber\\
&-&\frac{i}{2}I_{0imu}^{pqq}\Delta_{mn}^{qq}\Gamma_{njkl}^{qqqq}
\Delta_{lu}^{qq},\label{3dpqq}
\end{eqnarray} which can be represented graphically as

\begin{center}
\includegraphics[width=0.45\textwidth ]{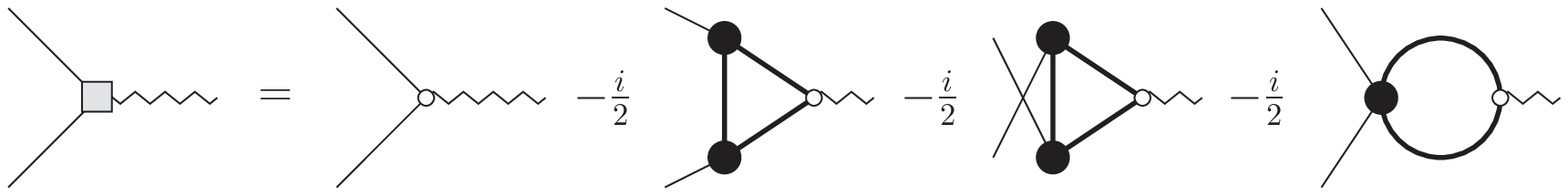}
\end{center}
We note in passing, that similar to the composite collective propagator and 
composite leg in the last section this "vertex" allows to write the graphs 
in the first line by a single loop graph. Yet, here this is then surely no 
explicit decomposition in terms of second order correlation functions anymore.

\section{Decomposition of proper Hamilton correlation functions}

While in the context of the canonical formulation it is entirely
possible to deduce the complete set of DSEs
directly from Eqs.\ (\ref{mge12}-\ref{sppdeqqqpr}) we will follow a
slightly less obvious path here. We proceed to give the diagrammatic
and analytic expressions  for the  proper propagators, typically the ones of
the first order formalism.  Subsequently,  we  will show that such
representations  can be encoded into the usual ones, 
\emph{i.e.}~Eqs.~(\ref{mge12}-\ref{sppdeqqqpr}), which then completes 
the proof of equivalence
between both types of derivations.

\subsection{The inverse propagators\label{sub:proper-hamilton}}

\begin{figure*}
\begin{center}
\includegraphics[width=\textwidth]{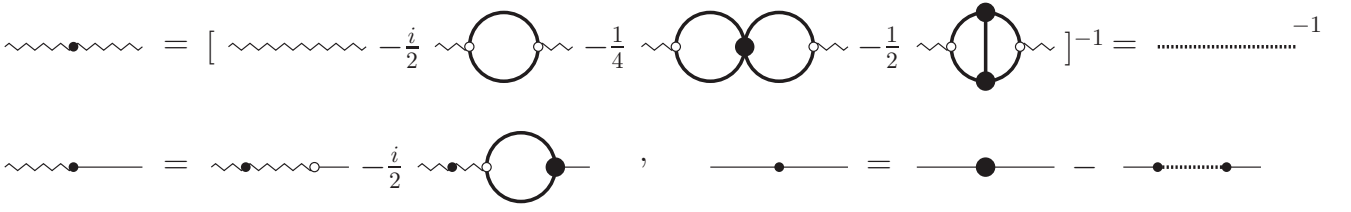}
\end{center}
\caption{Decomposition of the proper 2-point functions in the first order formalism in
terms of the usual correlation functions of the standard path integral
representation. \label{fig.proper}}
\end{figure*}

The result given by Eq.~(\ref{generalg}) allows to analyze  the
structure of the inverse propagator. By considering the Fourier transformation
of this equation we find that
\begin{eqnarray}
&\!\Gamma_{ij}^{pp}(k)\!\!&=-\left(\Delta^{pp}(k)+\Delta^{pq}(k)
\Gamma^{qq}(k)\Delta^{qp}(k)\right)_{ij}^{-1}\label{ftpp} \\
&\!\Gamma_{ij}^{pq}(k)\!\!&=\Gamma_{il}^{pp}(k)\Delta_{lm}^{pq}(k)
\Gamma_{mj}^{qq}(k) \\
&\!\Gamma_{ij}^{Hqq}(k)\!\!&=\Gamma_{ij}^{qq}(k)\!+\!\Gamma_{il}^{qq}(k)
\Delta_{lk}^{qp}(k)\Gamma_{km}^{pp}(k)\Delta_{mn}^{pq}(k) \Gamma_{nj}^{qq}(k).
\nonumber \\\label{fhais}
\end{eqnarray}
The presence of the second term in  Eq.\ (\ref{ftpp}) leads to the cancelation
of the corresponding term in Eq.\ (\ref{ppexpf}).    The complete analytic
result when inserting the corresponding expressions is rather lengthy and
therefore deferred to appendix \ref{app:exp-form}. It involves the inverse of
sums of  different terms. Graphically these expressions are represented in Fig.\
\ref{fig.proper}. For a clearer presentation the subsequent correlation
functions are represented in terms of the previous ones. Whereas the actual
momentum propagator can only be given as an inverse and thereby by an infinite
number of graphs, the "alternative propagator" $\mathcal{D}^{pp}$, that featured
prominently in the decomposition of Lagrange correlation functions, is given by
a finite number of graphs in terms of second order functions and differs from
the  ordinary  one by the absence of the connected pieces in Fig.\
\ref{fig.proper}.

For a field theory without a three-point interaction
vertex  involving the time derivative of the fields, one has
$\Gamma^{pp}=-\mathbb{I}$,  and Eq.\ (\ref{generalg}) reduces to
\begin{equation}
\mathbb{G}=\left(\begin{array}{cccc}
 -\mathbb{I} &  - ik_0 \mathbb{I} \qquad
 \\   &   \\  ik_0 \mathbb{I}  &\Gamma^{qq}-k_0^2\mathbb{I}\end{array}\right)
\label{generalgk}.
\end{equation}
This simplified expression holds for theories
like  QED and/or self-interacting $\phi^4-$theory, where the  vacuum
expectation value of the  momentum field  is completely determined
in terms of $q$, $\bar{p}_i=\dot{\bar{q}}_i$.

Actually, the expressions for $\Gamma^{pp}$,  $\Gamma^{qp}$ and
$\Gamma^{pq}$ in Eq.\ (\ref{generalgk}) encode a more general result
because they could be obtained  using Eq.\ (\ref{mge12}) and
Eq.\ (\ref{mge13}) without setting variables  to
zero. As a consequence they show that dressed proper functions
involving more than two external  momentum legs  are not present in
such theories. This means that the mass and the coupling
constant receive no contribution coming from the higher order
corrections besides the usual ones given in the standard framework. On the
other hand the absence of quantum corrections to $\Gamma^{pp},$
$\Gamma^{qp}$, and $\Gamma^{pq}$ means that only wave-function
renormalization contributes to these kinetic terms.

As in the present framework  the  quantum corrections to the propagators and
vertices depend in general on  temporal  derivatives  this shows that
within the standard formalism there are several pieces in the Effective
Action that depend on the time derivative of the averaged field. As a
consequence,
the canonical momentum fields defined on the level of the Effective Action
differ
from those given by the quantum average of $p$ since the involved limiting
processes do not commute.

\subsection{Recovering the first order DSEs}

Although the decomposition of proper Hamilton functions given above involves
inversions that cannot be omitted in terms of second order correlation functions
it is possible to transform these equations into a form where such matrix
inversions are eliminated. This is done by explicitly introducing first order
correlation functions on the right hand side again. As we will show in
this subsection, this leads precisely to the first order DSEs. We start from
Eq.\ (\ref{3dpqq}) and use it to express $\Gamma^{pp}$  as
\begin{equation}
\Gamma_{ij}^{pp}=-\left[\delta_{ij}-\frac{i}{2}I_{0ikl}^{pqq}\Delta_{km}^{qq}
\Delta_{ln}^{qq}\frac{\delta
\bar{p}_j}{\delta
\bar{q}_n\delta\bar{q}_m}\right]^{-1}.\label{newrpp}
\end{equation}
The multiplication of $-\left.\Gamma^{pp}\right.^{-1}$
from the left hand side allows to write the above equation as
\begin{eqnarray}
\left[\delta_{is}-\frac{i}{2}I_{0ikl}^{pqq}\left(\Delta_{kls}^{qqp}-
\Delta_{km}^{qq}\Delta_{ln}^{qq}\Gamma_{umn}^{qqq}\Delta_{us}^{qp}\right)
\right]\Gamma_{sj}^{pp}=-\delta_{ij},\nonumber
\end{eqnarray}
where Eq.\ (\ref{fopp1}) has been taken into account.
The introduction of a $\mathbb{I}$ as
$-\Delta^{qq}\Gamma^{qq}$ in the last term in the bracket and the use
of Eq.\ (\ref{fopp1}) makes it possible to express the above relation in
the following form
\begin{equation}
\Gamma_{ij}^{pp}=-\delta_{ij}+\frac{i}{2}I_{0ikl}^{pqq}
\left\{\Delta_{kls}^{qqp}\Gamma_{sj}^{pp}+\Delta_{kls}^{qqq}\Gamma_{sj}^{qp}
\right\},
\end{equation} which can be translated to
\begin{equation}
\Gamma_{ij}^{pp}=-\delta_{ij}-\frac{i}{2}I_{0ikl}^{pqq}\Delta_{kn}^{q\phi}
\Gamma_{njm}^{\phi
p\phi}\Delta_{ml}^{\phi q}. \label{sombra111}
\end{equation}
This presents the DSE for the proper momentum 2-point
correlation function.

It is possible to
obtain a similar equation for the $pq-$propagator:
\begin{eqnarray}
\Gamma_{ij}^{pq}&=&\left[\delta_{il}-\frac{i}{2}I_{0ikt}^{pqq}\Delta_{km}^{qq}
\Delta_{tn}^{qq}\frac{\delta^2
\bar{p}_l}{\delta
\bar{q}_n\delta\bar{q}_m}\right]^{-1} \nonumber\\ &&\quad \times 
\left(I_{0lj}^{pq} - \frac{i}{2}I_{0luk}^{pqq}\Delta_{um}^{qq}
\Gamma_{mjn}^{qqq}\Delta_{nk}^{qq}\right) . \label{sofr}
\end{eqnarray}
The
multiplication of $-\left.\Gamma^{pp}\right.^{-1}$ from the right hand
side allows to rewrite Eq.\ (\ref{sofr}) as
\begin{eqnarray}
\Gamma_{ij}^{pq}&=&I_{0ij}^{pq}-\frac{i}{2}I_{0iuk}^{pqq}\Delta_{um}^{qq}
\Gamma_{mnj}^{qqq}\Delta_{nk}^{qq}\nonumber\\&&+\frac{i}{2}I_{0iuk}^{pqq}
\Delta_{um}^{qq}
\Delta_{nk}^{qq} \frac{\delta
\bar{p}_l}{\delta
\bar{q}_m\delta\bar{q}_n}\Gamma_{lj}^{pq}.\label{sombra}
\end{eqnarray}
This can be rewritten in the following form
\begin{eqnarray}
\Gamma_{ij}^{pq}&=&I_{0ij}^{p
q}-\frac{i}{2}I_{0iuk}^{pqq}\left\{\Delta_{um}^{qq}\Gamma_{mnj}^{qqq}
\Delta_{nk}^{qq}\right.
\nonumber\\&& \quad \left. +\Delta_{um}^{qq}\Delta_{nk}^{qq}\Gamma_{mnl}^{qqq}
\Delta_{lv}^{qp}\Gamma_{vj}^{pq}-\Delta_{ukl}^{qqp}
\Gamma_{lj}^{pq}\right\}.\;\;\;\;\label{sombra1}
\end{eqnarray}
The  introduction of a $\mathbb{I}$ as $-\Delta^{qq}\Gamma^{qq}$ in
the first and second term inside the brackets allows to
write this as
\begin{eqnarray}
\Gamma_{ij}^{pq}&=&I_{0ij}^{pq}+\frac{i}{2}I_{0iuk}^{pqq}\left\{
\Delta_{ukl}^{qqq}
\left(\Gamma_{lj}^{qq}\!+\!\Gamma_{lm}^{qq}\Delta_{mn}^{qp}\Gamma_{nj}^{pq}
\right) \right. 
\nonumber\\&& \qquad \qquad \qquad \left. +
\Delta_{ukl}^{qqp}\Gamma_{lj}^{pq}\right\} .
\end{eqnarray}
By considering  Eq.\ (\ref{shrepr}) we get
\begin{eqnarray}
\Gamma_{ij}^{pq}&=&I_{0ij}^{p
q}+\frac{i}{2}I_{0iuk}^{pqq}\left\{\Delta_{ukl}^{qqq}\left.\Gamma_{lj}^{qq}
\right.^H+\Delta_{ukl}^{qqp}\Gamma_{lj}^{pq}\right\}
\end{eqnarray} 
which yields the result
\begin{equation}
\Gamma_{ij}^{pq}=I_{0ij}^{pq}-\frac{i}{2}I_{0ikl}^{pqq}\Delta_{kn}^{q\phi}
\Gamma_{njm}^{\phi
q\phi}\Delta_{ml}^{\phi q}.\label{sombra222}
\end{equation}

We remark that the last terms of Eqs.\ (\ref{sombra111})  and (\ref{sombra222})
include several combinations of fundamental and momentum fields.  The relations given by Eq.\
(\ref{sombra111}) and Eq.\ (\ref{sombra222}) are the DSEs of the proper
$pp$- and $pq$-propagators expressed in terms of the usual
elements of the canonical formulation.

It is straightforward  to prove that  Eq.\ (\ref{sombra111})
coincides with the corresponding equation  derived  using Eq.\ (\ref{mge12})
and Eq.\ (\ref{gqcm}). This is different for Eq.\ (\ref{sombra222}).
In fact,  as discussed previously, we could in principle calculate these
propagators from  Eq.\ (\ref{sppdeqqqpr}), nevertheless by considering the
action  Eq.\ (\ref{a}) and the relations between the bare elements we find
that the structure is more cumbersome than that given in Eq.\ (\ref{sombra222}).
Instead,  the latter one is in correspondence with Eq.\ (\ref{mge13}) via
the symmetry relation $\Gamma_{ij}^{pq}=\Gamma_{ji}^{qp}$ in case both
$\bar{q}$ and  $\bar{p}$ are bosonic fields.

The fact that we recover the standard first order propagator DSEs from the
decomposition of proper Hamilton correlation functions has its origin in the
equivalence
between the canonical and Lagrange equations of motion at the quantum level.
In appendix \ref{motion-equivalence} it is shown that whenever they describe the
same dynamical processes, the DSEs derived from them will be equivalent too.
However, as we argued in the last subsection, the classical canonical momentum fields
defined from the Effective Action and those given by the quantum average of $p$
are not the same. This means that the equations expressed in term of
$\bar{p}$ and $\bar{q}$ do not correspond to the classical canonical ones.

Based on this statement, the derivation of $\left.\Gamma_{ij}^{qq}\right.^H$
is considerably simpler using Eq.\ (\ref{sppdeqqq}) than via the procedure
performed in the last two cases.  Indeed, by considering Eqs.\ (\ref{gqcm})
and (\ref{a}) as well as the relations between the bare elements, we arrive at
\begin{eqnarray}
\frac{\delta
\Gamma^H}{\delta\bar{q}_i}&=&\left[I_{0il}^{qp}p_l+I_{0ijk}^{qqp}q_jp_k+
I_{0ij}^{qq}q_j+\frac{1}{2}I_{ijk}^{qqq}q_jq_k\right.\nonumber\\
&+&\left.\frac{1}{3!}I_{0ijkl}^{qqqq}q_jq_kq_l
\right]_{q\to \bar{q}+\frac{\hbar}{i}\Delta^{q\phi}
\frac{\delta}{\delta \bar{\phi}}} .\label{tdasd}
\end{eqnarray}

Now, in order to give  the explicit form of $\left.\Gamma_{ij}^{qq}\right.^H$,
we rewrite Eq.\ (\ref{tdasd}) as
\begin{eqnarray}
\frac{\delta\Gamma^H}{\delta\bar{q}_i}&=&I_{0ij}^{qp}\bar{p}_j
+I_{0ij}^{qq}\bar{q}_j+I_{0ijk}^{qqp}\bar{q}_j\bar{p}_k
+\frac{1}{2}I_{0ijk}^{qqq}\bar{q}_j\bar{q}_k\nonumber\\
&+&\frac{1}{3!}I_{0ijkl}^{qqqq}\bar{q}_j\bar{q}_k\bar{q}_l-iI_{0ijk}^{qqp}
\Delta_{jk}^{qp}[J]-\frac{i}{2}I_{0ijk}^{qqq}\Delta_{jk}^{qq}[J]\nonumber\\
&-&\frac{1}{6}I_{0ijkl}^{qqqq}\Delta_{jm}^{q\phi}[J]\Delta_{ku}^{q\phi}[J]
\Gamma_{mun}^{\phi\phi\phi}[\bar{q},\bar{p}]\Delta_{nl}^{\phi q}[J_q]
\nonumber\\
&-&\frac{i}{2}I_{0ijkl}^{qqqq}\bar{q}_j\Delta_{kl}^{qq}[J].
\end{eqnarray}

Taking the functional derivative with respect to $\bar{q}$ and setting the
vacuum expectation values of the fundamental fields to zero we get
\begin{eqnarray}
\Gamma_{ij}^H&=& I_{0ij}^{qq}-\frac{i}{2}I_{ijkl}^{qqqq}\Delta_{kl}^{qq}-
iI_{0jlm}^{qqp}\Delta_{lu}^{q\phi}\Gamma_{uin}^{\phi q\phi}\Delta_{nm}^{\phi p}
\nonumber\\
&-&\frac{i}{2}I_{0jlm}^{qqq}\Delta_{lu}^{q\phi}\Gamma_{uin}^{\phi q\phi}
\Delta_{nm}^{\phi q}-\frac{1}{2}I_{jlmn}^{qqqq}\Delta_{lh}^{q\phi}
\Gamma_{hfk}^{\phi\phi\phi}\Delta_{fy}^{\phi\phi}\Delta_{mu}^{q\phi}\nonumber\\
&\times&\Gamma_{uiy}^{\phi q\phi}\Delta_{kn}^{\phi q}-
\frac{1}{6}I_{jlmn}^{qqqq}\Delta_{lh}^{q\phi}\Delta_{mf}^{q\phi}
\Gamma_{hfik}^{\phi\phi q\phi}\Delta_{kn}^{\phi q}.\label{ph4vt}
\end{eqnarray}

The Fourier transformation of Eqs.\ (\ref{sombra111}), (\ref{sombra222}) and
(\ref{ph4vt}) yields finally the corresponding DSEs in momentum space
\begin{widetext}
\begin{eqnarray}
\Gamma_{ij}^{pp}(k)=-\delta_{ij}-
\frac{i}{2}\int\dbar\omega I_{0ikl}^{pqq}(k,\omega-k,-\omega)
\Delta_{kn}^{q\phi}(k-\omega)\Gamma_{njm}^{\phi
p\phi}(k-\omega,-k,\omega)\Delta_{ml}^{\phi q}(-\omega)
\end{eqnarray}
\begin{eqnarray}
\Gamma_{ij}^{pq}(k)=-ik_0\delta_{ij}- \frac{i}{2}\int\dbar\omega
I_{0ikl}^{pqq}(k,\omega-k,-\omega)
\Delta_{kn}^{q\phi}(k-\omega)\Gamma_{njm}^{\phi
p\phi}(k-\omega,-k,-\omega)\Delta_{ml}^{\phi q}(-\omega)
\end{eqnarray}
\begin{eqnarray}
\left.\Gamma_{ij}^{qq}\right.^H(k)&=&I_{0ij}^{qq}(k)-
\frac{i}{2}\int \dbar \omega
I_{0ijkl}^{qqqq}(k,-k,-\omega,\omega)\Delta_{kl}^{qq}(\omega)-i\int
\dbar\omega
I_{0ilm}^{qqp}(k,\omega-k,-\omega)\Delta_{lu}^{q\phi}(k-\omega)
\Gamma_{ujn}^{\phi q\phi}(k-\omega,-k,\omega)\nonumber\\
&\times&\Delta_{nm}^{\phi p}(-\omega)-\frac{i}{2}\int \dbar\omega
I_{0ilm}^{qqq}(k,\omega-k,-\omega)\Delta_{lu}^{q\phi}(k-\omega)
\Gamma_{ujm}^{\phi q \phi}(k-\omega,-k,\omega)\Delta_{nm}^{\phi
q}(-\omega)
\nonumber\\
&-&\frac{1}{2}\int\dbar\omega\dbar\mu
I_{0ilmn}^{qqqq}(k,-\mu-\omega,\omega-k,\mu)\nonumber
\Delta_{lh}^{q\phi}(\omega+\mu)\Gamma_{hfk}^{\phi\phi\phi}(\omega+\mu,-\omega,-\mu)
\Delta_{fy}^{\phi\phi}(\omega)\Delta_{mu}^{q\phi}(k-\omega)\nonumber\\
&\times&\Gamma_{ujy}^{\phi
q\phi}(k-\omega,-k,\omega)\Delta_{kn}^{\phi q} (\mu)-
\frac{1}{6}\int\dbar\omega\dbar\mu
I_{0ilmn}^{qqqq}(k,\omega-k-\mu,-\omega,\mu)
\Delta_{lh}^{q\phi}(k+\mu-\omega)\Delta_{mf}^{q\phi}(\omega)\nonumber\\&
\times&\Gamma_{hfjk}^{\phi\phi q \phi}(k+\mu-\omega,\omega,-k,-\mu)
\Delta_{kn}^{\phi q}(\mu)\label{granequ}
\end{eqnarray}
\begin{figure*}
\includegraphics[width=\textwidth]{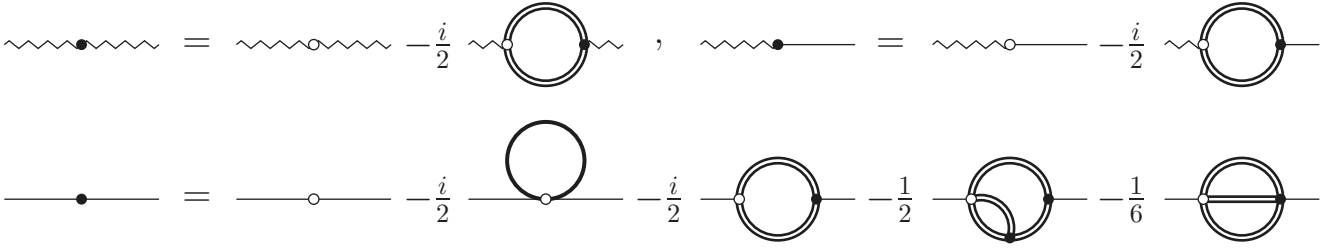}
\caption{Diagrammatical representation of the coupled system of
DSEs within the first order formalism. The double lines represent the matrix
propagator so that all possible graphs involving the individual propagators
arise which are compatible with the symmetries and the restriction that the 
$pqq$-vertex is the
only bare vertex involving momentum lines. In this form the formal equivalence
to the second order equation is entirely manifest.}\label{f1}
\end{figure*}
\end{widetext}
We show the graphical representation of the complete set of DSEs in Fig.\
\ref{f1} where for conciseness we use the matrix propagator which yields all
possible loop graphs involving physical vertices in accordance with the
symmetries of the action as a consequence of the implied summation over repeated
indices.

\section{Theories with auxiliary fields}

Before we come to the main application of the developed formalism in the
context of Coulomb gauge QCD, we will show in this section that the above
derived results also apply to theories involving auxiliary fields,  {\it i.e.} 
the typical treatment of such theories represent a special case of the 
discussion presented here. However, it is not the kinetic but the interaction
terms which are linearized. In the case of fermionic theories this is also
referred to as bosonization \cite{Hubbard:1959ub}.  The action is by
construction only quadratic in the auxiliary fields and lacks kinetic terms for
them. Therefore, as we detail below the above described analysis directly
applies and gives general relations between correlation functions in the
fundamental theory and the linearized form involving auxiliary fields.

We will illustrate this in the case of fermionic theories with
non-renormalizable, quartic interactions. Examples for this class of theories
are the Nambu$-$Jona-Lasinio model \cite{Nambu:1961tp} or the BCS theory of
superconductivity \cite{Bardeen:1957mv}. The functional integral is given by
\begin{equation}
\int \!\! D \psi^{\dagger} D\psi \exp \left( \frac{i}{\hbar}
\int d^4 x \left( \mathcal{L}_\psi\!+\!J_{\bar\psi}(x) \bar\psi(x)\!+\!
J_{\psi}(x) \psi(x) \right) \right)
\end{equation}
with the Lagrangian containing local quartic interactions
\begin{equation}
\mathcal{L_\psi}=\bar\psi(x) \left( i \!\! \not\!\partial-m_\psi \right)
\psi(x)-\sum_i g_\psi^i \left( \bar\psi(x) \Gamma_i \psi(x) \right)^2 .
\label{fermact}
\end{equation}
Here the $\Gamma_i$ are Dirac matrices. The linearization of the fermionic
interaction can be performed by formally introducing a one of the form
\begin{equation}
\mathbb{I} = \prod_i \!\! \int \!\! D \eta_i D\sigma_i \exp \left( i \!
\int \!\!d^4 x \sigma_i(x) \left( \eta_i(x) \!-\! \bar\psi(x) \Gamma_i \psi(x)
\right) \right)
\end{equation}
into the fermionic path integral, where this path integral over
$\sigma_i$ enforces a functional $\delta$-function that allows to rewrite
the non-linear fermionic interaction in terms of $\eta_i$. Integrating then
over the $\eta_i$ yields the path integral of the corresponding linear sigma
model
\begin{eqnarray}
&&\int \!\! D\psi^{\dagger} D \psi D \sigma \exp \left( \frac{i}{\hbar}
\int d^4 x \left( \mathcal{L}_\sigma \right. \right. \\
&& \qquad \qquad \qquad \left. \left. +\! J_{\bar\psi}(x) \bar\psi(x)\!+\!
J_{\psi}(x) \psi(x) \!+\! J^i_\sigma(x) \sigma_i(x) \right) \right) \nonumber
\end{eqnarray}
where we have introduced additional sources for the auxiliary fields $\sigma_i$.
After this bosonization procedure the Lagrangian of the corresponding
"linear $\sigma$-model" reads
\begin{equation}
\mathcal{L}_\sigma=\bar\psi(x) \Bigl( i \!\! \not\!\partial-m_\psi +
g_\sigma^i \Gamma_i \sigma_i(x) \Bigr) \psi(x)+\frac{m_\sigma^2}{2}
\sigma_i(x)^2 \label{lsm}
\end{equation}
where $g_\psi=g_\sigma^2/(2 m_\sigma^2)$.
Here, there is, in contrast to the first order formalism, by construction no
mixing between the fundamental and the auxiliary fields.

Analogous to Eq.\ (\ref{equivalenciaquantica}) the Effective Actions of the two
theories are again identical at vanishing sources $J_{\sigma}^i=0$
\begin{equation}
\Gamma_\psi[{\bar \psi}^\dagger,\bar \psi]=\Gamma_\sigma[{\bar \psi}^\dagger,
\bar \psi,\bar \sigma_i[{\bar \psi}^\dagger,\bar \psi]]
\end{equation}
where the auxiliary fields are implicitly given by
\begin{equation}
\frac{\delta \Gamma_\sigma}{\delta \bar \sigma_i}=-J_\sigma^i=0 .
\end{equation}
Via the chain rule of functional differentiation we obtain analogous expressions
for Eqs.\ (\ref{opstart})-(\ref{propder}) with $\bar p_i$ replaced by
$\bar \sigma_i$, respectively. Correspondingly, the same graphical rules apply.
With the simplification that there can be no mixing due to the different
statistics of the fields, the external legs are not composite and the arising
inverse 2-point function is the ordinary $\sigma$-propagator. Denoting as
before connected correlation functions that are 1PI with respect to the
fundamental fields and connected in the auxiliary fields as
{\em $\sigma$-connected}, this leads to the analogous general result:
{\quotation \em
A proper $n$-point function in the fundamental theory can be decomposed into
the sum of all $\sigma$-connected $n$-point functions in the linearized
theory.}

\bigskip

\noindent
In particular the decomposition of the proper 4-fermion vertex in the
fundamental theory Eq. (\ref{fermact}) reads
\begin{center}
\includegraphics[width=80mm]{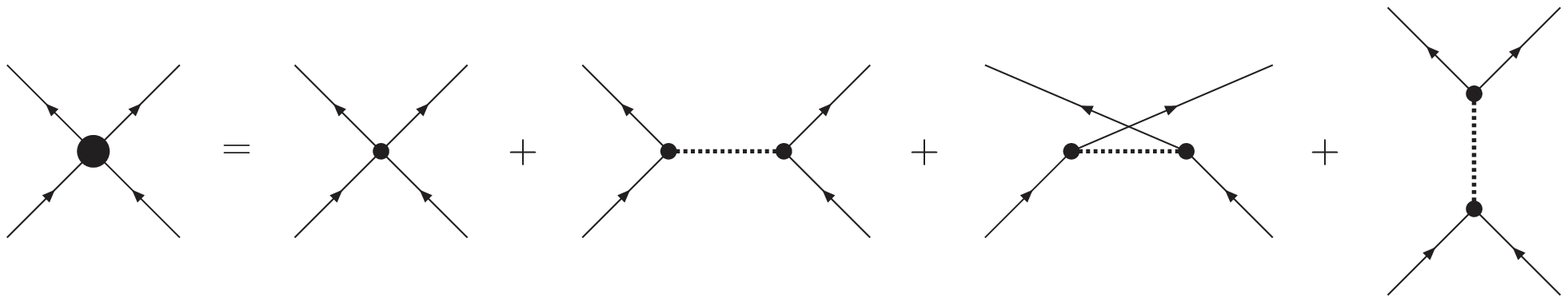}
\end{center}
where we represent in an analogous way the proper vertices in the fundamental
and linearized theory by large and small blobs and the ordinary
$\sigma$-propagator by the dotted line. This result confirms a well-known fact,
namely, that no double-counting  occurs in the linearized theory  although the
original fermion field and the auxiliary bosonic field are employed both. 
This redundancy in the description can also
be prevented from the outset by partial re-bosonization \cite{Gies:2002nw} in
the context of the functional renormalization group. In this approach
 the contribution of
the fundamental degrees of freedeom is entirely absorbed into the bosonized
interactions even at the level of the effective action.

Since the Lagrangian of the linearized theory is at most quadratic and lacks 
kinetic terms for the $\sigma$-fields by construction, these fields can be
trivially integrated out retaining their sources  at
this point. This leads to an analogous expression to Eq.\ (\ref{mge}) for the
correlation function of $n$ auxiliary fields
$\left< \sigma_{i_1}(x_1) \cdots \sigma_{i_n}(x_n) \right>$
after the sources are set to their vacuum expectation value.
Due to the absence of 3-point interactions and mixing, the decomposition of
the auxiliary $\sigma$-propagator is again simplified compared to the result
given in Fig.\ \ref{fig:mom-dec},
\begin{center}
\includegraphics[width=80mm]{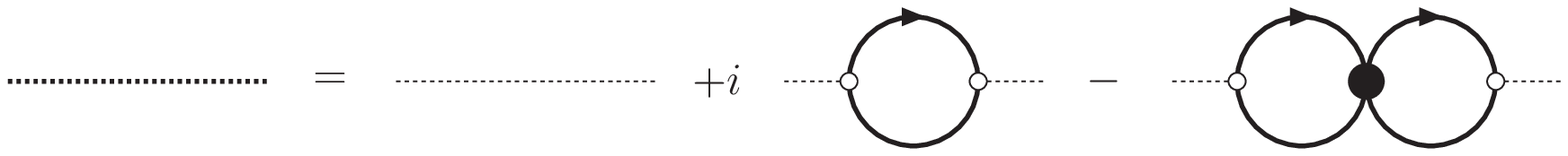}
\end{center}
where the different prefactors of the loop correction arise due to the fermionic
nature of the fields.
In contrast to the decomposition of the Hamilton propagators there is no
difference between proper and connected 2-point functions in the case of
the auxiliary field due to the absence of mixing, and the proper correlation
functions  is simply the inverse of the above equation.
This concludes the demonstration of the developed formalism to the case of
theories with auxiliary fields.

\section{Coulomb Gauge Yang-Mills theory \label{sec:coulomb}}

As detailed in the Introduction the main motivation for developing the 
presented formalism is given by the fact that the first order formalism might be
better suited for non-perturbative studies of Coulomb gauge Yang-Mills theory.
As a first step into this direction we will apply the formalism developed so far
to give the explicit relations between the    two-point correlation functions
of  the transversal and longitudinal  components  of the conjugate momentum to
the ones of the gauge field.

The starting point is the Hamiltonian density of Coulomb gauge Yang-Mills
theory:
\begin{equation}
\mathcal{H}_{\textrm{Coul}}=\mathcal{H}_{\rm YM}+\mathcal{H}_{\rm
A},
\end{equation} 
where $\mathcal{H}_{\rm YM}$ is the part of the
Hamiltonian density of pure Yang-Mills  theory
\begin{eqnarray}
\mathcal{H}_{\rm
YM}&=&\frac{1}{2}\left(\vec{p}^{a}\cdot\vec{p}^a+\vec{B}^a\cdot\vec{B}^a\right)
-\vec{p}^a\cdot
\vec{D}^{ab}\sigma^b\label{fpia}.\;\;
\end{eqnarray} 
Here, $B_i^a=\epsilon_{ijk}\left(\nabla_jA_k^a-
\frac{1}{2}gf^{abc}A_j^bA_k^c\right)$ 
represents   the chromomagnetic field, $\sigma^b$ the time-component of the
gluon field, whereas $\vec{D}^{ac}=\vec{\nabla}\delta^{ac}-gf^{abc}\vec{A}^b$ is
the covariant  derivative in the adjoint representation, with the structure
constants $f^{abc}$ of the color-group $SU(3)$, and $g$ denotes the gauge
coupling. On the other hand,
\begin{eqnarray}
\mathcal{H}_{\rm {A}}=\lambda^a\vec{\nabla}\cdot\vec{A}^a+\bar{c}^a\vec{\nabla}
\cdot\vec{D}^{ab}c^b
\end{eqnarray}
is the piece of the Hamiltonian density related  to gauge fixing
terms. Above $\vec{p}$ is the conjugate momentum of  the gauge
field $A^{\mu,a} \equiv(\vec{A}^a,\sigma^a),$   $\bar{c}$ and $c$ are
the Grassmann-valued Faddeev-Popov ghost fields introduced by fixing
the gauge, and  $\lambda^a$  is a ``colored" Lagrange multiplier field.
Note that the ``canonical action'' in the present case is given by
\begin{equation}
I_0=\int d^4x\left(\vec{p}^a\cdot\dot{\vec{A}^a}-\mathcal{H}_{\rm Coul}\right).
\end{equation} 
It is remarkable that the ghost sector of the Hamiltonian is fully disconnected 
from  $\vec{p}^a.$   By identification of $q=(\vec{A}^a,\sigma^a,\lambda^a)$ 
the above  ``action''  can be written as a functional Taylor expansion 
involving a pure bosonic piece like  Eq.~(\ref{can-action}) and one containing 
the ghost  field as  Eq. (\ref{fp2}). This fact allows therefore to use the 
general expressions  derived so far.

Except for the inverse, bare ghost two-point function
\begin{equation}
I_{0}^{ab(\bar{c}c)}=-\vec{k}^2\delta^{ab},
\end{equation} the remaining inverse tree level propagators of the theory  
are presented in Table \ref{tab:w1}. Note that $I_0$  involves four-vertices 
that have the following form in momentum space
\begin{eqnarray}
I_{0ij}^{abc (pA \sigma)}&=&-gf^{abc}\delta_{ij},\label{realvert}\\ 
I_{0i}^{abc (\bar{c}c A)}&=&i g k_{\bar{c} i}f^{abc},\\  
I_{0ijl}^{abc(AAA)}&=&igf^{abc}\left[\delta_{ij}(k_a-k_b)_l+
\delta_{jl}(k_b-k_c)_i\right.\nonumber\\ &&+\left.\delta_{li}(k_c-k_a)_j\right],\\
I_{0ijlm}^{abcd(AAAA)}
&=&-g^2\left\{\delta_{ij}\delta_{lm}\left[f^{ace}f^{bde}-f^{ade}f^{cbe}\right]
\right.\nonumber\\
&&+\delta_{il}\delta_{jm}\left[f^{abe}f^{cde}-f^{ade}f^{bce}\right]\nonumber\\
&&+\left.\delta_{im}\delta_{jl}\left[f^{ace}f^{dbe}-f^{abe}f^{cde}\right]
\right\},\qquad
\end{eqnarray}
with all momenta defined as incoming.

The decomposition of the  conjugate momentum  into  transverse and
longitudinal  parts,
$\vec{p}^a=\vec{\pi}^a-\vec{\nabla}\cdot\Omega^a$, makes it convenient
to  study the required complete cancellation of the energy
divergences \cite{Zwanziger:1998ez,Watson:2006yq,Watson:2007mz} that
emerge in a perturbative treatment of Coulomb gauge Yang-Mills
theory. Certainly, such a decomposition increases the number of fields
of the theory and leads to
\begin{eqnarray}
\tilde{I}_0&=&\int
d^4x\left(\vec{\pi}^a\cdot\dot{\vec{A}^a}-\vec{\nabla}\Omega^a\cdot
\dot{\vec{A}^a}+\frac{1}{2}\Omega^a\nabla^2\Omega^a\right.\nonumber\\
&-&\left.\mathcal{H}_{\rm
Coul}(\vec{p}\to\vec{\pi})-\tau^a\vec{\nabla}\cdot\vec{\pi}^a-\vec{\nabla}
\Omega^a\cdot\vec{D}^{ab}\sigma^b\right)\label{aanskdsfdvfsa}, \qquad \;\;
\end{eqnarray} 
where $\tau$ is a colored Langrange multiplier  field which appears via the 
transversality condition  of  $\vec{\pi}.$
In this context,  the general structure of the  connected two-point
functions  is  given  in Table \ref{tab:w2}. The latter  may be
found independent of any approximation but based  solely on the principles of
BRST invariance,  spatial and time-reversal   symmetries and
transversality properties of the vector propagators. Each dressing
function $\Lambda_{\phi\phi}$ is a dimensionless scalar function of
$k_0^2$ and $\vec{k}^2$ except the ghost propagator
\begin{equation}
\Delta^{ab(\bar{c}c)}=\delta^{ab} \frac{\Lambda^c(\vec{k}^2)}{\vec{k}^2}
\end{equation}
depending only on $\vec{k}^2$.
The tree-level propagators are obtained  when
\begin{equation}
\begin{array}{c}
\Lambda_{\Omega\Omega}=\Lambda_{\Omega\lambda}=0,\\
\Lambda_{AA}=\Lambda_{A\pi}=\Lambda_{\pi\pi}=\Lambda_{\sigma\sigma}=
\Lambda_{\sigma\Omega}=\Lambda_{\sigma\lambda}=\Lambda^c=1.
\end{array}
\end{equation} 
For a complete description, the reader is referred to 
ref.~\cite{Watson:2006yq}. 

\begin{figure*}
\begin{center}
\includegraphics[width=0.95\textwidth]{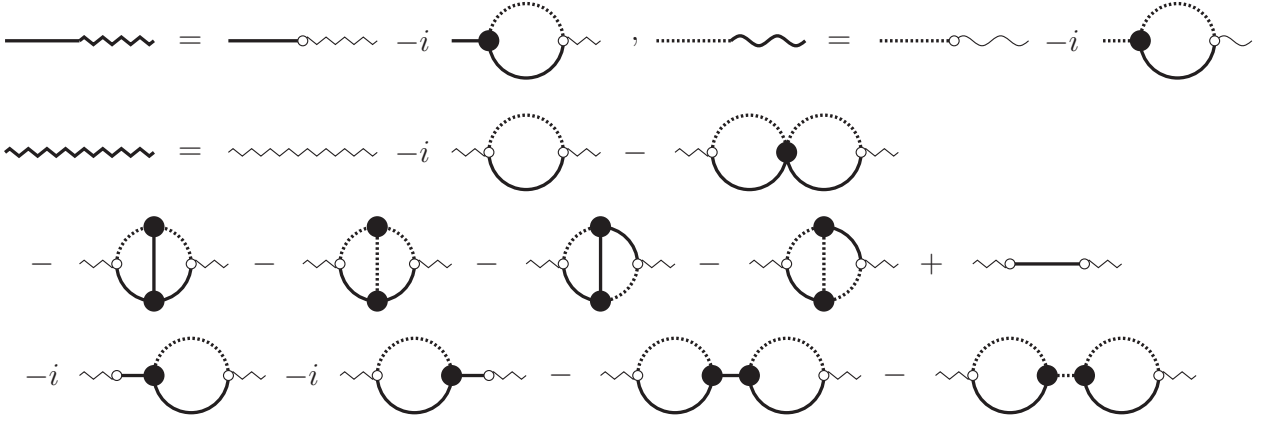}
\end{center}
\vspace*{-0.5cm}
\caption{Decomposition of the proper 2-point functions of Coulomb gauge QCD in 
the first order formalism in
terms of the corresponding correlation functions of the second order
representation. The spatial ($A$) and temporal ($\sigma$) gauge fields are 
represented by solid respectively dotted lines whereas the corresponding 
transverse  ($\pi$) and longitudinal ($\Omega$) momenta by zigzag respectively 
wavy lines. The equation for the longitudinal momentum propagator is identical 
to the one for the transverse component and given by replacing zigzag by wavy 
lines. \label{fig:coulomb-decomposition}}
\end{figure*}

The general results of the preceeding sections can in principle directly 
applied to Coulomb gauge QCD as well by decomposing the momentum field 
into its individual components. However, since the longitudinal momentum field 
features more complicated bare correlation functions that involve additional 
derivative operators we derive the corresponding expressions of subsection 
\ref{gen-Ham-con} once more taking now the general action (\ref{gac})
into account.  To this end we first have to relate the bare 
momentum correlation functions with the corresponding ones for the individual 
momentum components. We formally express the decomposition via a the operator 
$\mathcal{X}$ 
\begin{equation}
p_i=\pi_i+\partial_i\Omega=\pi_i+\mathcal{X}_{ij}\Omega_j\ \ \mathrm{and}\ \ 
\mathcal{X}_{ij}=\frac{\delta p_i}{\delta\Omega_j}=-\nabla_i\delta_{ij} .
\end{equation} 
Expanded in a local series in terms of the individual momentum components the 
general canonical action takes the form
\begin{eqnarray}
&&I_0[q,p]=I_{0ji}^{qp}\pi_iq_j+\frac{1}{2}I_{0ij}^{pp}\pi_i\pi_j+
\frac{1}{2}I_{0ijk}^{pqq}\pi_iq_jq_k\\
&&\quad+I_{0ji}^{qp}\mathcal{X}_{il}\Omega_l q_j\nonumber+
\frac{1}{2}\mathcal{X}_{il}I_{0ij}^{pp}\mathcal{X}_{jk}\Omega_l\Omega_k+ 
\frac{1}{2}I_{0ijk}^{pqq}\mathcal{X}_{il}\Omega_l q_jq_k\\
&&\quad+\frac{1}{2}I_{0ij}^{qq}q_iq_j +\frac{1}{3!}I_{0ijk}^{qqq}q_iq_jq_k+ 
\frac{1}{4!}I_{0ijkl}^{qqqq}q_iq_jq_kq_l \nonumber 
\label{can-action2}
\end{eqnarray} 
with  $I_{0ij}^{pp}=-\delta_{ij}.$  The above  expression yields 
for the transverse tree level correlators
\begin{equation}
I_{0ij}^{\pi q}=I_{0ij}^{pq}, \ \ I_{0ij}^{\pi \pi}=I_{0ij}^{pp}
\end{equation} 
and 
\begin{equation}
I_{0ijk}^{\pi qq}=\left.\frac{\delta}{\delta\pi_i}
\frac{\delta}{\delta q_j}\frac{\delta}{\delta q_k} I_0
\right\vert_{\pi,\Omega, q=0}=I_{0ijk}^{pqq},
\end{equation} 
as well as analogously for the longitudinal expressions
\begin{equation}
I_{0ij}^{\Omega q}=I_{0jl}^{qp}\mathcal{X}_{li},
\end{equation}
\begin{equation} 
I_{0ij}^{\Omega \Omega}=\mathcal{X}_{li}I_{0lk}^{pp}\mathcal{X}_{kj}=
-\mathcal{X}_{li}\delta_{lk}\mathcal{X}_{kj}=-\nabla_j^2\delta_{ji} 
\end{equation} 
and
\begin{equation}
I_{0ijk}^{\Omega qq}=\left.\frac{\delta}{\delta\Omega_i}
\frac{\delta}{\delta q_j}\frac{\delta}{\delta q_k} I_0
\right\vert_{\pi,\Omega, q=0}=-I_{0ijk}^{pqq}\nabla_i\delta_{ii^\prime} .
\end{equation} 
The Gaussian integration over $\pi$ yields 
\begin{eqnarray}
&&S[q,\Omega, J^\pi,J^\Omega,J^q]=\mathcal{S}_0+\frac{1}{2}J^\pi_iJ^\pi_i
+J^\pi_i\frac{\delta\mathcal{S}_0}{\delta \dot{q}_i} \nonumber \\
&&+I_{0ji}^{q\Omega}\Omega_iq_j+\frac{1}{2}\Omega_jI_{oij}^{\Omega\Omega}
\Omega_i+\frac{1}{2}I_{0ijk}^{\Omega qq}\Omega_i q_jq_k 
\nonumber \\
&&+J_i^{\Omega}\Omega_i
+J_i^{q}q_i 
\label{can-actionvar}.
\end{eqnarray}
where $\mathcal{S}_0$ is given by Eq. $(62)$ and $J^i$ are the corresponding 
sources associated to the momentum fields. Similarly, the subsequent Gaussian 
integration over $\Omega$ gives 
\begin{widetext}
\begin{eqnarray}
\tilde{S}[q,J^\pi,J^\Omega,J^q]&=&\mathcal{S}_0+\frac{1}{2}J^\pi_iJ^\pi_i+
J^\pi_i\frac{\delta\mathcal{S}_0}{\delta \dot{q}_i}+J_i^qq_i 
-\frac{1}{2}J^\Omega_i \left(I_{0}^{\Omega\Omega}\right)_{ij}^{-1}J^\Omega_i 
\!-\! J^\Omega_i\left(I_{0}^{\Omega\Omega}\right)_{ij}^{-1}\!
\left(\!I_{0uj}^{q\Omega}q_u+\frac{1}{2}I_{0juk}^{\Omega qq}q_uq_k\!\right) 
\nonumber \\
&-&\frac{1}{2}I_{0ji}^{q\Omega}\left(I_{0}^{\Omega\Omega}\right)_{il}^{-1}
I_{0ml}^{q\Omega}q_jq_m-\frac{1}{2}I_{0ji}^{q\Omega}\left(I_{0}^{\Omega\Omega}
\right)_{il}^{-1}I_{0lmu}^{\Omega qq }q_jq_mq_u
-\frac{1}{8}I_{0imu}^{\Omega qq }\left(I_{0}^{\Omega\Omega}\right)_{ij}^{-1}
I_{0jkl}^{\Omega qq }q_mq_uq_lq_k .
\nonumber \\
\label{can-actionvar2}
\end{eqnarray}
\end{widetext}
By collecting the terms of the same order in $q$ we obtain the bare vertices 
in the second order formalism  in terms of those that arise in the first order 
formalism. Thereby, the master equation for the momentum propagator 
Eq.~(\ref{fomp}) becomes in the case of the $\Omega$-field
\begin{eqnarray}
\Delta_{ij}^{\Omega\Omega}&=&\left(I_{0}^{\Omega\Omega}\right)_{il}^{-1}
\left[\delta_{lm}+i\left(I_{0ul}^{q\Omega}q_u+\frac{1}{2}I_{0lku}^{\Omega qq}
q_kq_u\right) \right. \label{eq:OmegaOmega} \\
&&\qquad\qquad \left. \times \left(I_{0sm}^{q\Omega}q_s+
\frac{1}{2}I_{0msd}^{\Omega qq}q_sq_d\right)\right]\left(I_{0}^{\Omega\Omega}
\right)_{mj}^{-1} \nonumber
\end{eqnarray}
whereas the corresponding mixed version Eq.~(\ref{fopp}) is given by
\begin{eqnarray}
\Delta_{ij}^{q\Omega}=-\Delta_{il}^{qq}\frac{\delta}{\delta q_l}
\left(I_{0um}^{q\Omega}q_u+\frac{1}{2}I_{mku}^{\Omega qq}q_kq_u\right)
\left(I_{0}^{\Omega\Omega}\right)_{mj}^{-1} . \quad\label{eq:Omegaq}
\end{eqnarray}
Here the $q$ fields in Eqs.~(\ref{eq:OmegaOmega}) and (\ref{eq:Omegaq}) 
again have to be replaced by
\begin{equation} 
q\to\bar{q}[J^q]+\frac{\hbar}{i}\Delta^{qq}[J^q]
\frac{\delta}{\delta q} .
\end{equation} 
The corresponding equations involving the $\pi$ field are identical to 
Eqs.~(\ref{fomp}) and (\ref{fopp}). Performing the same
algebraic steps as in sect.~\ref{sec.con-ham} yields the 
corresponding decomposition of the Hamilton propagators in Coulomb gauge QCD.
The result is displayed in diagrammatic form in 
Fig.~\ref{fig:coulomb-decomposition}, it constitutes 
the main result of this section. 
We have also checked the structure of these expressions by an 
explicit projection of the general equations on the corresponding 
momentum components in appendix \ref{app:projection}.

Next we will consider the other direction of the connection discussed in 
Sect.~\ref{sec:lagrange-decomposition}, the one which expresses proper Lagrange 
correlators in terms of Hamiltonian ones. This is interesting in the context  of
Coulomb gauge QCD, since in the Hamilton framework a complete proof of the
renormalizability of the theory seems possible due to explicit cancellations
ensured by powerful Ward identities \cite{Zwanziger:1998ez}. Here again it is
the particular definition of the  longitudinal momentum that complicates the
issue and does not allow to give these equations  explicitly.  Nevertheless, one
can easily convince oneself that these equations are given entirely by tree
graphs  that introduce no additional divergences. Therefore, once the 
renormalizability of the theory in the first order formalism is established, the
corresponding  connection immediately implies renormalizability of the theory
also in the Lagrange framework. Yet the cancellation mechanism of arising
divergences might be far from obvious in the latter framework and could
nevertheless prevent simple truncation schemes.

\begin{table}
\begin{tabular}{c|c c|c c }\hline\hline
$I_0$ & $A_j$ & $p_j$ & $\sigma$ & $\lambda$  \\
\hline\rule[-2.4ex]{0ex}{5.5ex}
$A_i\ \ $&$-\mathbb{T}\vec{k}^2$&
$ik_0\mathbb{I}$&$-k_0\vec{k}$&
$-i\vec{k}$\\
$p_i\ \ $&$-ik_0\mathbb{I}$&
$-\mathbb{I}$&$-i\vec{k}$&0\\
$\sigma\ \ $&0&$i\vec{k}$&$0$&
$0$\\
$\lambda\ \ $&$i\vec{k}$&0&0&
0\\
\hline\hline
\end{tabular}
\caption{\label{tab:w1} Tree level proper two point functions 
(without color factors) in momentum space.}
\end{table}

\begin{table}
\begin{tabular}{c|c c|c c|c c}\hline\hline
&&&&&\\ $\mathcal{W}^H$ & $A_j$ & $\pi_j$ & $\sigma$ & $\Omega$ & 
$\lambda$ & $\tau$ \\
\hline\rule[-2.4ex]{0ex}{5.5ex}
$A_i$&$\mathbb{T}
\frac{-\Lambda_{AA}}{(k_0^2-\vec{k}^2)}$&
$\mathbb{T}
\frac{-ik^0\Lambda_{A\pi}}{(k_0^2-\vec{k}^2)}$&0&0&
$\frac{-ik_i}{\vec{k}^2}$&0\\
$\pi_i$&$\mathbb{T}
\frac{ik^0\Lambda_{A\pi}}{(k_0^2-\vec{k}^2)}$&
$\mathbb{T}
\frac{-\vec{k}^2\Lambda_{\pi\pi}}{(k_0^2-\vec{k}^2)}$&0&0&0&
$\frac{-ik_i}{\vec{k}^2}$\\
$\sigma$&0&0&$\frac{-\Lambda_{\sigma\sigma}}{\vec{k}^2}$&
$\frac{\Lambda_{\sigma\Omega}}{\vec{k}^2}$&
$\frac{-ik^0\Lambda_{\sigma\lambda}}{\vec{k}^2}$&0\\
$\Omega$&0&0&$\frac{\Lambda_{\sigma\Omega}}{\vec{k}^2}$&
$\frac{\Lambda_{\Omega\Omega}}{\vec{k}^2}$&
$\frac{-ik^0\Lambda_{\lambda}}{\vec{k}^2}$&$\frac{-1}{\vec{k}^2}$\\
$\lambda$&$\frac{ik_j}{\vec{k}^2}$&0&$\frac{ik^0\Lambda_{\sigma\lambda}}{\vec{k}^2}$&
$\frac{ik^0\Lambda_{\Omega\lambda}}{\vec{k}^2}$&0&0\\
$\tau$&0&$\frac{ik_j}{\vec{k}^2}$&0&$\frac{-1}{\vec{k}^2}$&0&0\\
\hline\hline
\end{tabular}
\caption{\label{tab:w2}General form of propagators in momentum space.  The
global color factor $\delta^{ab}$ has been extracted.  All unknown functions
$\Lambda_{\phi\phi}$ are dimensionless, scalar functions of $k_0^2$ and $\vec{k}^2$.
Here $\mathbb{T}_{ij}
=\mathbb{I}_{ij}-k_i k_j/\vec{k}^2$ is the 
transverse projector in momentum space.}
\end{table}

Next we consider the functional symmetry identity.
The canonical action $I_0$ is  BRST-invariant, {\it i.e.} invariant under
\begin{eqnarray}
\delta\vec{A}^a&=&\frac{1}{g}\vec{D}^{ac}c^c\delta\lambda,\ \ 
\delta\sigma^a=-\frac{1}{g}D^{0ac}c^c\delta\lambda,\nonumber\\
\delta\bar{c}^a&=&\frac{1}{g}\lambda^a \delta\lambda,\ \ 
\delta c^a=-\frac{1}{2}f^{abc}c^bc^c\delta\lambda,\\
\delta\vec{p}^a&=&f^{abc}c^b\delta\lambda
\left[(1-\alpha)\vec{p}^c-\alpha \vec{E}^c\right],\ \ 
\delta\lambda^a=0. \nonumber
\end{eqnarray}
 Here $\delta\lambda$ is a Grassmannian infinitesimal parameter, 
whereas $D^{0ac}=\delta^{ac}\partial_0+gf^{abc}\sigma^b.$ On the other hand 
$\vec{E}^a=-\vec{\nabla}\sigma^a-D^{0ac}\vec{A}^c$ is the chromoelectric field. 
Note in addition that  $\alpha$ is some color-singlet constant which in general 
could be some function of position.

By considering the above transformation the Slavnov-Taylor identity  reads
\begin{eqnarray}
0&=&\int\! \!\mathcal{D}[\phi]\! \!\int \! \! d^4x
\left\{-\frac{1}{g}\rho^aD^{0ab}c^b+
\frac{1}{g}\vec{j}^a\cdot\vec{D}^{ab}c^b-\frac{1}{g}\lambda^a\eta^a\right.
\nonumber\\
&-&\left.\frac{1}{2}f^{abc}\bar{\eta}^ac^bc^c+f^{abc}c^b\left[(1-\alpha)
\vec{p}^c+\alpha\vec{E}^c\right]\cdot\vec{J}_p\right\}\nonumber\\
&\times&\exp\left\{i I_0+iI_s\right\},
\end{eqnarray} 
where
\begin{equation}
I_s=\int d^4x\left\{\rho^a\sigma^a+\vec{j}^a\cdot\vec{A}^a+
\bar{c}^a\eta^a+\bar{\eta}^ac^a+\vec{p}^{a}\cdot\vec{J}_p^a\right\}\nonumber.
\end{equation}

Employing a procedure analogous to the one used in Sect. \ref{sec:srifs}
we obtain
\begin{eqnarray}
0&=&\int\! \!\mathcal{D}[\tilde{\phi}]\! \!\int \! \! d^4x
\left\{-\frac{1}{g}\rho^aD^{0ab}c^b+\frac{1}{g}\vec{j}^a\cdot\vec{D}^{ab}c^b
-\frac{1}{g}\lambda^a\eta^a\right.\nonumber\\&-&\left.\frac{1}{2}f^{abc}
\bar{\eta}^ac^bc^c+f^{abc}c^b\left[(1-\alpha)\vec{J}_p^c-\vec{E}^c\right]
\cdot\vec{J}_p\right\}\nonumber\\&\times&\exp\left\{i S_0+iS_s\right\},
\end{eqnarray} 
where $\mathcal{D}[\tilde{\phi}]$ denotes the 
remaining  integration measure of the   fields $c, \bar{c}, \vec{A}$ and 
$\sigma$. Expressed in the field strength tensor
$F_{\mu\nu}^a=\partial_\mu A_{\nu}^a-\partial_\nu 
A_{\mu}^a+gf^{abc}A_{\mu}^bA_{\nu}^c$  one has
\begin{eqnarray}
S_{0}&=&\int d^4x \left\{-\frac{1}{4}F^{a\mu\nu}F_{\mu\nu}^a-
\mathcal{H}_{\rm{A}}\right\} \\
S_s&=&\int d^4x\left\{\rho^a\sigma^a+\vec{j}^a\cdot\vec{A}^a+\bar{c}^a\eta^a
+\bar{\eta}^ac^a-\vec{J}_p^a\cdot\vec{E}^a\right.\nonumber\\&+&\left.
\frac{1}{2}\vec{J}_p^a\cdot\vec{J}_p^a\right\}.
\end{eqnarray}
\begin{widetext}
In addition we decompose the source 
$\vec{J}^p_a=\vec{J^\pi_a}-\frac{\vec{\nabla}}{-\nabla^2}{J^\Omega_a}$
into transversal and longitudinal components.  
Employing this decomposoition the Slavnov-Taylor identity can be written as
\begin{eqnarray}
0&=&\int\mathcal{D}[\tilde{\phi}]\int
d^4x\left\{-\frac{1}{g}\rho^aD^{0ab}c^b+
\frac{1}{g}\vec{j}^a\cdot\vec{D}^{ab}c^b-
\frac{1}{g}\lambda^a\eta^a\right.\\&-&\left.
\frac{1}{2}f^{abc}\bar{\eta}^ac^bc^c
+f^{abc}c^b(1-\alpha)\left[\vec{J}^\pi_c\cdot\vec{J}^\pi_a+\frac{\vec{\nabla}}
{-\nabla^2}J^\Omega_c
\frac{\vec{\nabla}}{-\nabla^2}J^\Omega_a\right]\right.
-\left.f^{abc}c^b\left[\vec{E}_T^c\cdot\vec{J}^\pi_a-\vec{E}_L^c
\frac{\vec{\nabla}}{-\nabla^2}J^\Omega_a\right]\right\}\exp\left\{i
S_0+iS_s\right\},\nonumber
\end{eqnarray} 
\end{widetext}
where
\begin{equation}
E_T^{ci}=T^{ij}E^{ic}\ \  \textrm{and}\ \  E_L^{ci}=
\frac{\partial^i\partial^j}{\nabla^2} E^{jc}.
\end{equation}
This completes the application of the developed formalism to Coulomb gauge
Yang-Mills theory.

\section{Summary and Outlook}

We conclude  with our main result that given a quantum field theory
in the context of the first order formalism it is possible to decompose all
Green's functions in terms of those obtained from the second order formalism
and vice versa. Whereas proper Lagrange correlation functions
are given explicitly to all orders and involve only tree graphs involving
dressed Hamilton correlation functions, the decomposition of Hamilton
$n$-point functions involves loop graphs of loop order $n$. Although the
structure of the latter equations seems to be somewhat cumbersome, they are
still more compact and simple than the usual DSEs within the Hamilton formalism.
We have discussed the connection between the Hamilton and the Lagrange
formalism for a general quantum field theory and illustrated the detailed
structure of the arising relations in the important case of a generic
four-dimensional renormalizable field theory.

In accordance with the obtained equations we have argued that in theories
where the quantum average of the momentum fields is completely determined as
$\bar{p}=\dot{\bar{q}}$ the  proper $p-$propagators receives quantum
corrections only via wave-function renormalization.
Additionally, it has been  shown that the canonical
momentum fields which can be defined from the Effective Action and those given
by the quantum average of $p$ are in general different.

As a demonstration of the general nature of the presented
formalism we also showed
 that the results obtained in this paper can also be applied
to the case of theories involving auxiliary fields from a linearization of the
interaction part of the action. This yields general relations between
correlation functions in the fundamental theory and the linearized form
involving auxiliary fields. A major difference is that in this case the
fundamental and the auxiliary fields do not mix which simplifies the connection
considerably.

Clearly, the determination of Green's functions of  Yang-Mills theories in the
first  order formalism is considerably more complicated as this case, or even
when compared to scalar or Abelian gauge theories.
The presence of a coupling between the gauge fields and their time derivative
as well as the subtleties of gauge fixing present major complications.
In particular the renormalizability still poses a major challenge. 
The presented general connection of all Green functions in the two formulations 
implies that no additional divergences arise when going from one to 
the other. However, what still needs to be shown is that the theory is 
renormalizable in both formulations once a proof is given for either of them. 
On the other hand, renormalizing the theory is probably simpler in the first 
order formalism where energy divergences explicitly cancel. 
The obtained connections should then help to renormalize the second order 
theory where numerical calculations are far simpler. A detailed analysis of 
the renormalizability in Coulomb gauge Yang-Mills theory will therefore be 
presented in a forthcoming publication.

\begin{acknowledgments}
It is our pleasure to thank Peter Watson for helpful discussions.
S.~V.-C. is supported by the  Doktoratskolleg
``Hadrons in Vacuum, Nuclei and Stars'' of the Austrian science fund
(FWF) under contract W1203-N08.
K.~S. acknowledges support from the FWF under contract M979-N16, and
R.~A. from the German research foundation (DFG) under contract AL 279/5-2.
\end{acknowledgments}


\appendix

\section{$p-$integration of $\mathcal{Z}[J].$\label{app:p-int}}

In this appendix we perform the integration over the momentum fields in the
vacuum-vacuum transition amplitude in presence of the classical
sources. In order to do this, we assume a general quadratic Hamiltonian
density given by Eq.(\ref{Hamiltoniand1}). In order to evaluate the Gaussian
functional integral over the
momentum fields we first consider a finite dimensional
generating functional
\begin{equation}
\mathfrak{Z}[J]=\mathcal{N}_J\int_{-\infty}^{\infty}
\prod_{j=1}^{N-1} \! \! d^4q^j
\prod_{j=0}^{N}\! \! d^4p^j
\exp\left[\frac{i}{\hbar}(\tilde{I}_0-\mathcal{C}+J_q^iq^i)\right].
\end{equation}
with the momentum field dependent part of the action
\begin{equation}
\tilde{I}_0=-\frac{1}{2}p^i\mathcal{A}_{ij}p^j+\left(\dot{q}^i-
\mathcal{B}^i+J_p^i\right)p^i .
\end{equation}
By completing the squares in the above expression we arrive at
\begin{eqnarray} 
\tilde{I}_0&=&\frac{1}{2}\pi^i\left( - \mathcal{A}_{ij} \right) \pi^j
\nonumber \\ &+&
\frac{1}{2}\left[\dot{q}^i-\mathcal{B}^i+J_p^i\right]\mathcal{A}_{ij}^{-1}
\left[ \dot{q}^j-\mathcal{B}^j+J_p^j\right], 
\end{eqnarray} 
where
$\pi^i\equiv p^i-\mathcal{A}^{-1}_{ij}\left(\dot{q}^j-
\mathcal{B}^j+J_p^j\right)$.
By considering the change of the integration variables from $p^i$ to $\pi^i$
we obtain
\begin{equation}
\mathfrak{Z}[J]=\tilde{\mathcal{N}}_J\int\prod_{i=1}^{N-1}d^4q^j
\exp\left[\frac{i}{\hbar}\tilde{\mathcal{S}}[q,J]\right] \label{wqe}
\end{equation} 
with
\begin{equation}
\tilde{\mathcal{N}}_J=\mathcal{N}_J\int\prod_{i=0}^{N-1} d^4\pi^j
\exp\left[-\frac{i}{2}\pi^i\mathcal{A}^{ij}\pi^j\right]
\end{equation} 
and
\begin{eqnarray}
\tilde{S}[q,J]\! &=& \!\mathcal{S}_0[q]
\\ &+&
\frac{1}{2}\!J_p^i
\mathcal{A}^{-1}_{ij}[q]J_p^j+J_q^i q^i+J_p^i\mathcal{A}^{-1}_{ij}[q]
\!\left(\dot{q}^j\!-\!\mathcal{B}^j[q]\right).\nonumber  \label{gac}
\end{eqnarray}
Here,  the action $\mathcal{S}_0$  can be obtained by considering
the limit of $\tilde{\mathcal{S}}[q,J]$ when $J\to0$. Explicitly it reads
\begin{equation}
\mathcal{S}_0[q]=\frac{1}{2}\dot{q}^i\mathcal{A}^{-1}_{ij}[q]\dot{q}^j-
\dot{q}^i\mathcal{A}^{-1}_{ij}[q]\mathcal{B}^j[q]-\int
d^4x \mathcal{V}[q],
\end{equation}
where
\begin{equation}
\mathcal{V}[q]=-\frac{1}{2}\mathcal{B}^i[q]\mathcal{A}^{-1}_{ij}[q]\mathcal{B}^j[q]+
\mathcal{C}[q].
\end{equation}
The canonical momentum fields defined as
\begin{equation}
p_n^\mathrm{can}\equiv\frac{\delta\mathcal{S}_0}{\delta\dot{q}_i}=
\mathcal{A}^{-1}_{ij}[q]\left(\dot{q}^j-\mathcal{B}^j[q]\right)\label{cvbr}
\end{equation}
allow us to identify and substitute the last term in $\mathcal{S}[q,J]$
by its respective definition, and to eventually obtain the desired form of
Eq.\ (\ref{actionLF}).

\bigskip

\section{Proof of the general form of the decomposition of proper Lagrange
correlation functions}

In this appendix we prove the statement made in subsection
\ref{sub:proper-decomposition} that the replacement rules precisely generate
the $p$-connected correlators. As usual for a statement over the integers this
is done by induction. The statement is trivially fulfilled in the case
$n=3$ given explicitly before. Now let us assume it is fulfilled for all
integers $\leq n$ and show that this implies its validity for $n+1$.
In the $p$-connected diagrams with $n$ legs we single out the proper vertex
we started the iteration with. This vertex can be connected to other connected
clusters so that all arising graphs have the general form
\begin{center}
\includegraphics[width=0.4\textwidth]{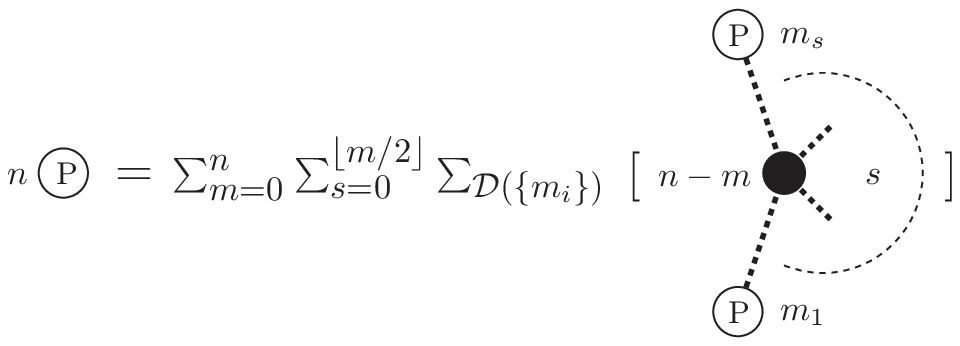}
\end{center}
Here we suppress all external legs and only give their number next to the
corresponding vertex. The sum over $s$ counts the connected clusters attached
to the considered proper vertex which in addition has $n-m$ external legs and
the implicitly given sum labeled $\mathcal{D}(\{m_i\})$ runs over all
ways to distribute the remaining $m$ external legs to the $s$
indistinguishable connected vertices (taking into account that a vertex has at
least 3 legs).

Next we consider the attachment of an additional external leg via the rules
given in Fig. \ref{fig:rules}, applied in all possible ways.
The above representation then goes over to a $n+1$-point function of the
following form
\begin{widetext}
\begin{center}
\includegraphics[width=0.8\textwidth]{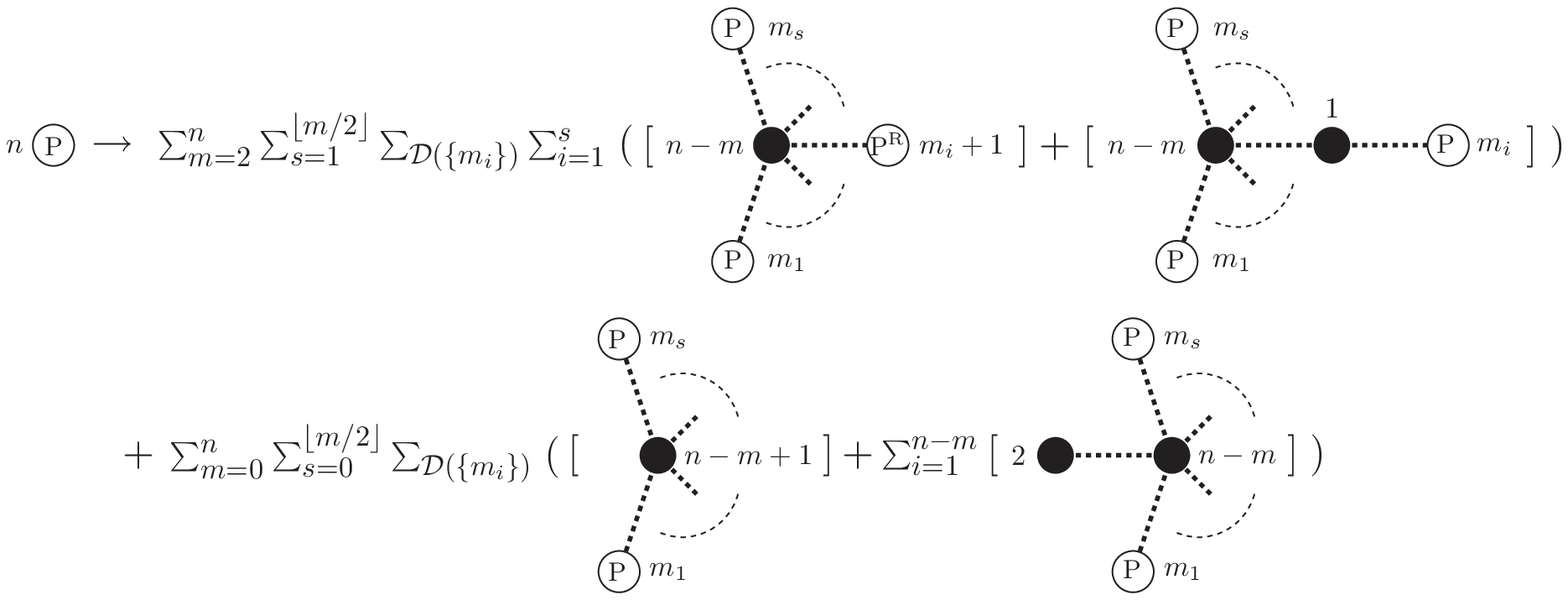}
\end{center}
\end{widetext}
Here the connected cluster with $n+1$ legs in the first class of graphs is
labeled by the additional index $R$ (for reduced) since one of the legs is
connected to the $p$-propagator instead to a composite external leg.
The derivative of this propagator is explicitly present via the second class
of graphs. When the new proper vertex is absorbed in the connected cluster,
according to the induction assumption ($m_i+1 < n$) these two terms together
precisely yield the full $p$-connected vertex with $m+1$ external legs and
without the index $R$. Since each connected cluster had before already at
least two external legs, together with the new one there are now at least
three. Terms where besides the new external leg there is only one other
external leg at a connected cluster are explicitly given by the fourth
class of graphs involving a new "connected cluster" that consists only of the
proper 3-point vertex. Finally, the general expression at order $n+1$ contains
also graphs where the new external leg is attached to the proper vertex itself,
 given by the third class of graphs. Altogether, the sum of the different
 classes of graphs precisely yields all necessary graphs at order $n+1$,
 which completes the proof.

\section{Explicit form of the decomposition of proper Hamilton correlation
functions\label{app:exp-form}}

In this Appendix we give the complete results for the decomposition of the
proper 2-point function in the Hamilton formalism in terms of those in the
Lagrange formalism. We represent the appearing operator inverses in terms of
Schwinger proper time integrals. The momentum correlation function is given as
\begin{widetext}
\begin{eqnarray}
\Gamma_{ij}^{pp}(k)&=&-\int_0^\infty
d\alpha\exp\left\{-\alpha\left[\delta_{ij}-\frac{i}{2}\int
\dbar\omega
I_{0ikl}^{pqq}(k,\omega-k,-\omega)I_{0nmj}^{qqp}(\omega,k-\omega,-k)
\Delta_{km}^{qq}(\omega-k)\Delta_{ln}^{qq}(\omega)\nonumber\right.\right.\\
&-&\frac{1}{4}\int\dbar\omega\dbar
\mu
I_{0ikl}^{pqq}(k,\mu-k,-\mu)I_{0nmj}^{qqp}(-\omega,\omega+k,-k)
\Delta_{ku}^{qq}(\mu-k)\Delta_{lx}^{qq}(-\mu)\Delta_{ny}^{qq}(-\omega)
\Gamma_{yxuz}^{qqqq}(\omega,\mu,k-\mu,-\omega-k)\nonumber\\
&\times&\Delta_{zm}^{qq}(-\omega-k)-\frac{1}{2}\int\dbar\omega\dbar
\mu
I_{0ikl}^{pqq}(k,\omega-k,-\omega)I_{0nmj}^{qqp}(\mu+k,-\mu,-k)
\Delta_{ku}^{qq}(\omega-k)\Gamma_{puq}^{qqq}(-k-\mu,k-\omega,\omega+\mu)
\nonumber\\
&\times&\left.\left.\Delta_{lx}^{qq}(-\omega)\Delta_{np}^{qq}(\mu+k)
\Delta_{qy}^{qq}(\omega+\mu)\Gamma_{yxz}^{qqq}(-\omega-\mu,\omega,\mu)
\Delta_{zm}^{qq}(\mu)
\right]\right\}.
\end{eqnarray}
whereas the mixed correlation function reads
\begin{eqnarray}
\Gamma_{ij}^{pq}&=&\int_0^\infty
d\alpha\exp\left\{-\alpha\left[\delta_{il}-\frac{i}{2}\int
\dbar\omega
I_{0iks}^{pqq}(k,\omega-k,-\omega)I_{0nml}^{qqp}(\omega,k-\omega,-k)
\Delta_{km}^{qq}(\omega-k)\Delta_{sn}^{qq}(\omega)\nonumber\right.\right.\\
&-&\frac{1}{4}\int\dbar\omega\dbar
\mu
I_{0iks}^{pqq}(k,\mu-k,-\mu)I_{0nml}^{qqp}(-\omega,\omega+k,-k)
\Delta_{ku}^{qq}(\mu-k)\Delta_{sx}^{qq}(-\mu)\Delta_{ny}^{qq}(-\omega)
\Gamma_{yxuz}^{qqqq}(\omega,\mu,k-\mu,-\omega-k)\nonumber\\
&\times&\Delta_{zm}^{qq}(-\omega-k)-\frac{1}{2}\int\dbar\omega\dbar
\mu
I_{0iks}^{pqq}(k,\omega-k,-\omega)I_{0nml}^{qqp}(\mu+k,-\mu,-k)
\Delta_{ku}^{qq}(\omega-k)\Gamma_{puq}^{qqq}(-k-\mu,k-\omega,\omega+\mu)
\nonumber\\
&\times&\left.\left.\Delta_{sx}^{qq}(-\omega)\Delta_{np}^{qq}(\mu+k)
\Delta_{qy}^{qq}(\omega+\mu)\Gamma_{yxz}^{qqq}(-\omega-\mu,\omega,\mu)
\Delta_{zm}^{qq}(\mu)
\right]\right\}\nonumber\\
&\times&
\left\{-ik_0\delta_{lj}-
\frac{i}{2}\int\dbar\sigma I_{0fhl}^{qqp}(-\sigma-k,\sigma,k) 
\Delta_{fg}^{qq}(\sigma+k)\Gamma_{gjw}^{qqq}(\sigma+k,-k,-\sigma)
\Delta_{wh}^{qq}(\sigma)\right\}.\label{sofr1}
\end{eqnarray}
The fundamental field correlation function is finally given by the
lengthy expression
\begin{eqnarray}
\left.\Gamma_{ij}^{qq}\right.^H(k)&=&\Gamma_{ij}^{qq}(k)-
\left\{ik_0\delta_{it}-
\frac{i}{2}\int\dbar\rho I_{0rvt}^{qqp}(k-\rho,\rho,-k)
\Delta_{rb}^{qq}(\rho-k)\Delta_{ev}^{qq}(\rho) 
\Gamma_{bie}^{qqq}(\rho-k,k,-\rho)\right\}\nonumber\\
&\times&\int_0^\infty
d\alpha\exp\left\{-\alpha\left[\delta_{tl}-\frac{i}{2}\int
\dbar\omega
I_{0tks}^{pqq}(k,\omega-k,-\omega)I_{0nml}^{qqp}(\omega,k-\omega,-k)
\Delta_{km}^{qq}(\omega-k)\Delta_{sn}^{qq}(\omega)\nonumber\right.\right.\\
&-&\frac{1}{4}\int\dbar\omega\dbar
\mu
I_{0tks}^{pqq}(k,\mu-k,-\mu)I_{0nml}^{qqp}(-\omega,\omega+k,-k)
\Delta_{ku}^{qq}(\mu-k)\Delta_{sx}^{qq}(-\mu)\Delta_{ny}^{qq}(-\omega)
\Gamma_{yxuz}^{qqqq}(\omega,\mu,k-\mu,-\omega-k)\nonumber\\
&\times&\Delta_{zm}^{qq}(-\omega-k)-\frac{1}{2}\int\dbar\omega\dbar
\mu
I_{0tks}^{pqq}(k,\omega-k,-\omega)I_{0nml}^{qqp}(\mu+k,-\mu,-k)
\Delta_{ku}^{qq}(\omega-k)\Gamma_{puq}^{qqq}(-k-\mu,k-\omega,\omega+\mu)
\nonumber\\
&\times&\left.\left.\Delta_{sx}^{qq}(-\omega)\Delta_{np}^{qq}(\mu+k)
\Delta_{qy}^{qq}(\omega+\mu)\Gamma_{yxz}^{qqq}(-\omega-\mu,\omega,\mu)
\Delta_{zm}^{qq}(\mu)
\right]\right\} \nonumber\\
&\times&\left\{-ik_0\delta_{lj}-
\frac{i}{2}\int\dbar\sigma I_{0fhl}^{qqp}(-\sigma-k,\sigma,k) 
\Delta_{fg}^{qq}(\sigma+k)\Gamma_{gjw}^{qqq}(\sigma+k,-k,-\sigma)
\Delta_{wh}^{qq}(\sigma)\right\}.\label{ph42}
\end{eqnarray}
\end{widetext}

\section{First order Dyson-Schwinger equations \label{motion-equivalence}}

In this appendix we derive the DSEs in the first order formalism directly. In
order to do this we compute  Eq.\ (\ref{sdev}) with the canonical action given
in Eq.\ (\ref{can-action}) 
\begin{eqnarray}\
\frac{\delta\Gamma^H}{\delta \bar{p}_i}
&=&\left.\frac{\delta I_0}{\delta p_i}
\right\vert_{\phi\to \bar{\phi}+
\frac{\hbar}{i}\Delta^{\phi\phi}\frac{\delta}{\delta \bar{\phi}}}\nonumber\\
&=&\left[-p_i+I_{0li}^{qp}q_l+\frac{1}{2}I_{0lki}^{qqp}q_lq_k
\right]_{\phi_\to \bar{\phi}+\frac{\hbar}{i}\Delta^{\phi\phi}
\frac{\delta}{\delta \bar{\phi}}}.\label{mty}
\end{eqnarray}
We can identify the sum of the last two terms  inside the bracket as the
quantum canonical momentum fields, which allows to write  the above relation as
Eq.\ (\ref{mge11}). \\
The substitution of Eq. (\ref{mty}) in Eq. (\ref{sdeqqq}), allows to write the
latter as
\begin{equation}
\frac{\delta \Gamma^H}{\delta\bar{q}_i}
=\left[\frac{\delta\mathcal{S}_0}{\delta q_i}\!+\!
\left(p_m\!-\!\frac{\delta\mathcal{S}_0}{\delta \dot{q}_m}\right)
\!\frac{\delta^2\mathcal{S}_0 }{\delta q_i\delta \dot{q}_m}
\right]_{\phi\to \bar{\phi}+\frac{\hbar}{i}\Delta^{\phi\phi}
\frac{\delta}{\delta \bar{\phi}}}\label{gfhs}
\end{equation} where
\begin{eqnarray}
\frac{\delta\mathcal{S}_0}{\delta q_i}&=& \mathcal{S}_{0ij}^{qq}q_j
+\frac{1}{2}\mathcal{S}_{0ijk}^{qqq}q_jq_k
+\frac{1}{3!}\mathcal{S}_{0ijkl}^{qqqq}q_jq_kq_l\nonumber\\
&+& I_{0jl}^{qp}I_{0li}^{pq}q_j+I_{0ij}^{qq}q_j
+I_{0ijm}^{qqp}I_{0mk}^{pq}q_jq_k\nonumber\\
&+&\frac{1}{2}I_{0jkm}^{qqp}I_{0mi}^{pq}q_jq_k+
\frac{1}{2}I_{0ijk}^{qqq}q_jq_k+\frac{1}{3!}I_{0ijkl}^{qqqq}q_jq_kq_l
\nonumber\\&+&\frac{1}{2}I_{0ijm}^{qqp}I_{0mkl}^{ppqq}q_jq_kq_l\label{rtyy}
\end{eqnarray}
and
\begin{eqnarray}
\left(p_m-\frac{\delta\mathcal{S}_0}{\delta
\dot{q}_m}\right)\frac{\delta^2\mathcal{S}_0 }{\delta q_i\delta
\dot{q}_m}&=&
I_{0im}^{qp}p_m+I_{0ijm}^{qqp}p_mq_j\nonumber\\
&-&I_{0jl}^{qp}I_{0li}^{pq}q_j-I_{0ijm}^{qqp}I_{0mk}^{pq}q_jq_k\nonumber\\
&-&\frac{1}{2}I_{0jkm}^{qqp}I_{0mi}^{pq}q_jq_k-\frac{1}{2}I_{0ijm}^{qqp}
\nonumber\\
&\times&I_{0mkl}^{ppq}q_jq_kq_l.\label{rty}
\end{eqnarray}
In the last two equations we have used  the elementary decomposition of the
bare elements and Eq. (\ref{gqcm}). Plugging Eqs.\ (\ref{rtyy}) and
(\ref{rty}) into Eq. (\ref{gfhs}) we arrive at the DSE (\ref{tdasd}).

\section{Explicit projection on the individual momentum 
components\label{app:projection}}

Defining the transverse projector $\textrm{T}_{ij}=\delta_{ij}-
\partial_i\partial_j/\nabla^2$ one has $\pi_{i}^a=\textrm{T}_{ij}p_{j}^a$  
and $\Omega^a=\frac{\vec{\nabla}}{-\nabla^2}\cdot\vec{p}^a$. As no 
time derivative appears one can obtain $\Delta_{ij}^{A\pi},$ 
$\Delta_{ij}^{\pi\pi},$
$\Delta_{ij}^{\sigma\Omega},$ and $\Delta_{ij}^{\Omega\Omega}$
by projecting $p$ in the two-point functions $\Delta_{ij}^{Ap}$, 
$\Delta_{ij}^{\sigma p}$
and $\Delta_{ij}^{pp}$.

For instance to obtain $\Delta_{ij}^{A\pi}$  we decode the  result
given in Eq. (\ref{qppp}).  Let us  first identify  the fundamental
field $q$ with index $i$ with $A_i^a.$   We then expand the
remaining sums  over the fields involved in the $q$'s considering  all
possible combinations which generate the  bare vertex
$I_{0ijk}^{(p\sigma A)}$. The bosonic symmetry of the latter allows
to write
\begin{equation}
\Delta_{ij}^{Ap}=\Delta_{il}^{AA}\left(I_{0lj}^{
Ap}-iI_{0ukj}^{\sigma A p}\Delta_{um}^{\sigma\sigma}
\Gamma_{mln}^{\sigma A A}\Delta_{nk}^{AA}
\right).\label{qpppvrie}
\end{equation}
Here no internal propagator like $\Delta_{ij}^{A\lambda}$ 
appears since no proper vertex functions with $\lambda-$derivative exist. 
The relation between this function and $\Delta_{ij}^{A\pi}$ is given by
\begin{eqnarray}
\Delta_{ij}^{A\pi}&=&\textrm{T}^{jr}\Delta_{ir}^{Ap}\nonumber \\
&=&\textrm{T}^{rj}\Delta_{il}^{AA}\left(I_{0lr}^{
Ap}-iI_{0ukr}^{\sigma A p}\Delta_{um}^{\sigma\sigma}
\Gamma_{mln}^{\sigma A A}\Delta_{nk}^{AA}\right) . \qquad
\label{lachacha}
\end{eqnarray}  
Obviously the case corresponding to 
$\Delta_{ij}^{\sigma\Omega}$ is obtained by replacing in Eq. (\ref{qpppvrie}) 
$\vec{A}\to\sigma$ where appropriate and considering the relation
\begin{eqnarray}
\Delta_{ij}^{\sigma\Omega}&=&\frac{\partial^j}{\nabla^2}
\Delta_{ij}^{\sigma p}\nonumber\\
&=&\frac{\partial^j}{\nabla^2}\Delta_{il}^{\sigma\sigma}\left(I_{0lj}^{
\sigma p}-iI_{0ukj}^{\sigma A p}\Delta_{um}^{\sigma\sigma}
\Gamma_{mln}^{\sigma \sigma A}\Delta_{nk}^{AA}\right).\quad\;\;
\label{lachachacha}
\end{eqnarray}  
Note that  the sum over $r$ in Eq. (\ref{lachacha})  
and over $j$ in Eq. (\ref{lachachacha}) must be  understood  over the 
discrete spatial indices only.

Clearly, by demanding a similar procedure we obtain that
\begin{equation}
\Delta_{ij}^{\pi\pi}=\textrm{T}^{ii^\prime}\textrm{T}^{jj^\prime}
\Delta_{i^\prime j^\prime}^{pp}\ \  \textrm{and}\ \ 
\Delta_{ij}^{\Omega\Omega}=\frac{\partial^{i^\prime}}{\nabla^2}
\frac{\partial^{j^\prime}}{\nabla^2}\Delta_{i^\prime j^\prime}^{pp} 
\end{equation} 
with
\begin{eqnarray}
&&\Delta_{ij}^{pp}=\delta_{ij}-\Delta_{im}^{pq}
\Gamma_{ml}^{qq}\Delta_{lj}^{qp}-iI_{0ikl}^{pA\sigma}
\Delta_{km}^{AA}\Delta_{ln}^{\sigma\sigma}I_{0nmj}^{\sigma A p}\nonumber\\
&&-I_{0ikl}^{pA\sigma}
\Delta_{ku}^{AA}\Delta_{lx}^{\sigma\sigma}\Delta_{ny}^{\sigma\sigma}
\Gamma_{yxuz}^{\sigma\sigma AA}
\Delta_{zm}^{AA}I_{0nmj}^{\sigma Ap}\nonumber\\&&
-I_{0ikl}^{pA\sigma}\Delta_{ku}^{AA}
\Delta_{lx}^{\sigma\sigma}\Delta_{np}^{\sigma\sigma}\Gamma_{puq}^{\sigma Aq}
\Delta_{qy}^{qq}\Gamma_{yxz}^{q\sigma A}\Delta_{zm}^{AA}I_{0nmj}^{\sigma Ap}
\nonumber\\ 
&&-I_{0ikl}^{pA\sigma}\Delta_{ku}^{AA}
\Delta_{lx}^{\sigma\sigma}\Delta_{np}^{AA}\Gamma_{puq}^{AAq}\Delta_{qy}^{qq}
\Gamma_{yxz}^{q\sigma\sigma}\Delta_{zm}^{\sigma\sigma}I_{0nmj}^{\sigma Ap}\;.
\qquad\quad
\label{ppexpf8}
\end{eqnarray} 
Note that the sum over all field components $q_i$ in the 
internal progators leads to multiple possibilities.

The Fourier transformation of these expressions read therefore
\begin{widetext}
\begin{eqnarray}
\Delta_{ij}^{ab (A\pi)}(k)&=&
\Delta_{il}^{ar(AA)}(k)\left(ik_0\textrm{T}_{lj}(\vec{k})\delta^{rb}- 
i\textrm{T}_{jv}(\vec{k})\int \dbar\omega I_{0uv}^{cdb(\sigma A p)}
(k-\omega,\omega,-k)\Delta_{um}^{de(AA)}(\omega-k)\Gamma_{ml}^{erh(AA\sigma)}
(\omega-k,k,-\omega)\right.\nonumber\\ &\times&\left.\Delta^{hc(\sigma\sigma)}
(\omega)\right),\label{sadadvcg}\\
\Delta^{ab (\sigma\Omega)}(k)&=&
\Delta^{al(\sigma\sigma)}(k)\left(-\delta^{lb}- \frac{k^j}{\vec{k}^2}
\int \dbar\omega I_{0uj}^{cdb(\sigma A p)}(k-\omega,\omega,-k)
\Delta_{um}^{de(AA)}(\omega-k)
\Gamma_{m}^{elh(A\sigma\sigma)}(\omega-k,k,-\omega)\right.\nonumber\\ 
&\times&\left.\Delta^{hc(\sigma\sigma)}(\omega)\right),\label{sadadvcgdfgh}\\
\Delta_{ij}^{ab(\pi\pi)}(k)&=&
\mathrm{T}_{ij}(\vec{k})\delta^{ab}-\Delta_{im}^{ae(\pi A)}(k)
\Gamma_{ml}^{ec(AA)}(k)
\Delta_{lj}^{cb(A\pi)}(k)-\mathrm{T}_{ii^\prime}(\vec{k})
\mathrm{T}_{ij^\prime}(\vec{k})\left\{i\int
\dbar\omega
I_{0i^\prime k}^{acd(p A \sigma)}(k,\omega-k,-\omega)\right.\nonumber\\ 
&\times& \Delta_{km}^{cf(AA)}(\omega-k)
\Delta^{de(\sigma\sigma)}(\omega)I_{0mj^\prime}^{efb (\sigma A p)}
(\omega,k-\omega,-k)+\int\dbar\omega\,\dbar \mu 
I_{0i^\prime k}^{acd (p A \sigma)}(k,\mu-k,-\mu)
\Delta_{ku}^{c n (AA)}(\mu-k)\nonumber\\
&\times&\Delta^{d f(\sigma\sigma)}(-\mu)\Delta^{e y(\sigma\sigma)}(-\omega)
\Gamma_{uz}^{yfnl(\sigma\sigma A A)}(\omega,\mu,k-\mu,-\omega-k)
\Delta_{zm}^{lf(AA)}(-\omega-k)I_{0mj^\prime}^{efb(\sigma A p)}
(-\omega,\omega+k,-k)\nonumber\\&+&\int\dbar\omega\,\dbar
\mu
I_{0i^\prime k}^{acd (p A\sigma)}(k,\omega-k,-\omega)
\Delta_{ku}^{c l(AA)}(\omega-k)\Gamma_{up}^{v l h(\sigma A q)}
(-k-\mu,k-\omega,\omega+\mu)
\Delta^{d x(\sigma\sigma)}(-\omega)\nonumber\\
&\times&\Delta^{e v(\sigma\sigma)}(\mu+k)
\Delta_{py}^{hs(qq)}(\omega+\mu)\Gamma_{yz}^{sxt(q\sigma A)}
(-\omega-\mu,\omega,\mu)
\Delta_{zm}^{tf(AA)}(\mu)I_{0mj^\prime}^{efb(\sigma Ap)}(\mu+k,-\mu,-k)
\nonumber\\&+&\int\dbar\omega\,\dbar
\mu
I_{0i^\prime k}^{acd (p A\sigma)}(k,\omega-k,-\omega)
\Delta_{ku}^{c l(AA)}(\omega-k)
\Gamma_{zup}^{v l h(A A q)}(-k-\mu,k-\omega,\omega+\mu)
\Delta_{zm}^{ev(AA)}(\mu+k)\nonumber\\
&\times&\Delta^{dx(\sigma\sigma)}(-\omega)
\Delta_{py}^{hs(qq)}(\omega+\mu)\Gamma_{y}^{sxt(q\sigma \sigma)}
(-\omega-\mu,\omega,\mu)
\Delta^{tf(\sigma\sigma)}(\mu)I_{0mj^\prime}^{efb(\sigma Ap)}
(\mu+k,-\mu,-k),
\\
\Delta^{ab(\Omega\Omega)}(k)&=&\frac{1}{\vec{k}^2}\delta^{ab}
-\Delta_{im}^{ae(\Omega \sigma)}(k)\Gamma_{ml}^{ec(\sigma\sigma)}(k)
\Delta_{lj}^{cb(\sigma\Omega)}(k)- \frac{k^ik^j}{\vec{k}^4}\left\{i\int
\dbar\omega
I_{0ik}^{acd(p A \sigma)}(k,\omega-k,-\omega)\right.\nonumber\\ 
&\times& \Delta_{km}^{cf(AA)}(\omega-k)\Delta^{de(\sigma\sigma)}
(\omega)I_{0mj}^{efb (\sigma A p)}(\omega,k-\omega,-k)
+\int\dbar\omega\,\dbar \mu I_{0i k}^{acd (p A \sigma)}(k,\mu-k,-\mu)
\Delta_{ku}^{c n (AA)}(\mu-k)\nonumber\\
&\times&\Delta^{d f(\sigma\sigma)}(-\mu)\Delta^{e y(\sigma\sigma)}(-\omega)
\Gamma_{uz}^{yfnl(\sigma\sigma A A)}(\omega,\mu,k-\mu,-\omega-k)
\Delta_{zm}^{lf(AA)}(-\omega-k)I_{0mj}^{efb(\sigma A p)}(-\omega,\omega+k,-k)
\nonumber\\&+&\int\dbar\omega\,\dbar
\mu
I_{0i k}^{acd (p A\sigma)}(k,\omega-k,-\omega)
\Delta_{ku}^{c l(AA)}(\omega-k)\Gamma_{up}^{v l h(\sigma A q)}
(-k-\mu,k-\omega,\omega+\mu)
\Delta^{d x(\sigma\sigma)}(-\omega)\nonumber\\
&\times&\Delta^{e v(\sigma\sigma)}(\mu+k)
\Delta_{py}^{hs(qq)}(\omega+\mu)\Gamma_{yz}^{sxt(q\sigma A)}(-\omega-\mu,\omega,\mu)
\Delta_{zm}^{tf(AA)}(\mu)I_{0mj}^{efb(\sigma Ap)}(\mu+k,-\mu,-k)
\nonumber\\&+&\int\dbar\omega\,\dbar
\mu
I_{0i k}^{acd (p A\sigma)}(k,\omega-k,-\omega)
\Delta_{ku}^{c l(AA)}(\omega-k)\Gamma_{zup}^{v l h(A A q)}
(-k-\mu,k-\omega,\omega+\mu)\Delta_{zm}^{ev(AA)}(\mu+k)\nonumber\\
&\times&\Delta^{dx(\sigma\sigma)}(-\omega)
\Delta_{py}^{hs(qq)}(\omega+\mu)\Gamma_{y}^{sxt(q\sigma \sigma)}
(-\omega-\mu,\omega,\mu)
\Delta^{tf(\sigma\sigma)}(\mu)I_{0mj}^{efb(\sigma Ap)}(\mu+k,-\mu,-k) \; .
\end{eqnarray}
\end{widetext}
The bare Greens functions involving the full momentum field can now be 
expressed by the corresponding expression for the individual components 
via the explicit expressions given in section \ref{sec:coulomb} which 
results in the equations given in diagrammatic form in 
Fig.~\ref{fig:coulomb-decomposition}.

\end{document}